\documentclass[11pt]{article}
\usepackage{amsmath,amsfonts,amssymb,fullpage,xcolor,soul,natbib,url,mathtools,booktabs}
\usepackage{graphicx,subcaption}
\usepackage{caption}
\def\dconv{\smash{\mathop{\longrightarrow}\limits^d}}     

\renewcommand\tilde{\widetilde}

\newtheorem{lemma}{Lemma}
\newtheorem{proposition}{Proposition}

\newtheorem{remark}{Remark}

\def\cov{\text{cov}}

\newcommand{\bpf}{\begin{proof}}
\newcommand{\epf}{\end{proof}}
\newcommand{\red}{\textcolor{red}}

\newcommand{\bbr}{{\mathbb R}}
\newcommand{\cO}[1]{\mathcal{O}(#1)}
\newcommand{\cip}{\overset{p}\to}

\begin{document}
\bibliographystyle{econometrica}
\title{Time Series Estimation of the Dynamic Effects of Disaster-Type  Shocks}

\author{Richard Davis\thanks{Department of Statistics, Columbia University, 1255 Amsterdam Avenue, MC4689, New York, NY 10027.
Email: rdavis@stat.columbia.edu} \and Serena Ng\thanks{Columbia University and NBER, 420 W. 118 St. MC 3308, New York, NY 10027. Email: serena.ng@columbia.edu \newline  We thank an anonymous referee and an Associate editor,  the Co-editor (Torben Andersen),  seminar participants at the Federal Reserve Bank of Philadelphia, and the 2022 Italian Workshop of Econometrics and Empirical Economics for many helpful comments. The first and second authors acknowledge financial support from the National Science Foundation under grants DMS-2015379 and SES-2018369, respectively.}}
\date{\today\bigskip}
\maketitle
\begin{abstract}
This paper provides three results for SVARs under the assumption that the primitive shocks are mutually independent. First,  a framework is proposed to accommodate a disaster-type variable with infinite variance into a SVAR.  We show that the  least squares estimates of the SVAR  are consistent but have non-standard asymptotics. Second, the disaster shock  is  identified as the  component with the largest kurtosis and whose impact effect is negative. An estimator that is robust to infinite variance is used to recover  the mutually independent components. Third, an independence test on the residuals pre-whitened by the Choleski decomposition is proposed to  test the restrictions imposed on a SVAR. The  test   can be applied whether the data have fat or thin tails, and to  over as well as  exactly identified  models.  Three applications are considered. In the first, the independence test is  used  to shed light on the conflicting evidence regarding the role of uncertainty in economic fluctuations. In the second,  disaster shocks are shown to have  short term  economic impact arising mostly from feedback dynamics. The third  uses  the framework to study the dynamic effects  of economic shocks post-covid.

\end{abstract}


\setcounter{page}{0}
\thispagestyle{empty}
\baselineskip=18.0pt
\newpage
\section{Introduction}
The novel coronavirus (\textsc{covid-19)} outbreak has drawn  attention to the modeling of rare events such as pandemics and natural disasters. How do we estimate the dynamic effects of disaster type shocks  on economic variables? How do we  estimate the dynamic effects of  economic shocks when the data are contaminated by rare events that do not have economic origins? Should  measures of disasters be modeled as exogenous?  A difficulty  in  predicting the occurrence of disasters and designing polices to mitigate their  impact is that there are few  such data points even over a long span.  After all, the CDC has only documented four influenza pandemics in the U.S. with deaths in excess of  100,000 over  a 120 year period starting in 1900.\footnote{These are the Spanish flu in 1918 (675,000 US deaths), the H2N2 virus in 1957-58 (116,000 US deaths), H3N2 virus in 1968 (100,000 US deaths), the H1N1 virus in 2009 (12,500 US deaths). Source \url{https://www.cdc.gov/flu/pandemic-resources/basics/past-pandemics.html}.}  For natural disasters, the 12,000 deaths from the Galveston  hurricane of 1900 remains a record,  with the 1200 deaths from Katrina coming in a distant second in terms of casualties.  Worldwide, only seven earthquakes since 1500 were  larger than 9 in magnitude,\footnote{Source: \url{https://en.wikipedia.org/wiki/Lists_of_earthquakes#Largest_earthquakes_by_magnitude}.} and  September 11 was the only terror attack on U.S. soil   with more than 300 deaths, let alone 3000. Nonetheless, when a rare disaster strikes, it strikes in a ferocious manner as \textsc{covid-19} reminds us. Though these events  have been intensely studied on a case by case basis, it is also of interest to study these events over a long time span.\footnote{For a review of methodologies used, see \citet{bds:19}.}  We  apply standard time series methodology to analyze the dynamic effects of rare events by modeling these events as being driven by  heavy-tailed  shocks. 

To fix ideas, consider Figure \ref{fig:cddd} which plots the real cost of 258  natural disasters over the period 1980:1-2019:12, augmented to include  9/11.\footnote{The series combines data from the National Oceanic and Atmospheric Administriation and the Insurance Information Institute  as explained in \citet{lmn3-PP:21}.}   The  series  is dominated by  a few events with  Hurricane Katrina in August 2005 being the largest, accounting for  9.2\% of total cost. This is followed by the four weeks in the summer of 2017 when  Hurricane Harvey contributed  7\%  in August, while Hurricanes  Irma and Maria in September created a combined cost of  8\%. These are followed by 9/11 in 2001 and superstorm Sandy in October 2012,  each contributing to about  5\% of total costs. Another measure of the cost of disasters is the number of lives lost. This series, while not plotted to conserve space, has spikes that are even more extreme.  Over the same time period,  49\% of disaster-related deaths can be attributed  to  Hurricanes Maria/Irma, 9/11, and Hurricane Katrina, with the heat wave of 1980 coming in fourth.  Both series have features of a heavy-tailed process, and we will subsequently use sample kurtosis as evidence of tail heaviness. 

 Heavy-tailed data   pack a lot of information in a few observations. Because of its large variability,  the dynamic  effects of  disaster shocks should in principle be consistently estimable.  Indeed, if all variables in a multivariate system have heavy tails, we show below that  the least squares estimator   will  converge at a fast rate of  $(\frac{T}{\ln T})^{1/\alpha}$ where  $\alpha$ is the  index of the heavy-tailed shock and $T$ is the sample size.  Though the distribution theory is a bit nonstandard, the  regression framework is the same as the standard case when  all variables have light tails.  But  while many macroeconomic time series have excess kurtosis, they do not fit the characterization of  heavy tails.   For example,  unemployment and industrial production  have kurtosis of less than 10, while the  disaster series shown in Figure \ref{fig:cddd} has kurtosis in excess of 70,  and the estimated tail index of approximately one suggests  a distribution with infinite variance and possibly infinite mean.\footnote{The  method often used to estimate the tail index is due to \citet{hill:75}.}  \citet{beare-toda:20} analyzed \textsc{covid-19} cases across US counties and finds that the right tail of the distribution has a Pareto exponent close to one.   This motivates a new multivariate framework in which   finite and infinite variance shocks co-exist in such a way that the economic variables can be affected by heavy-tailed shocks but not dominated by them. 

 Our  point of departure  is that the $n$  primitive shocks $u=(u_1,\ldots,u_n)$  are assumed to be  mutually independent, a condition stronger than the commonly used assumption of mutual orthogonality that is no longer meaningful when one of the shocks has infinite variance. We develop a  HL (`heavy-light')  framework in which  the coefficient estimates on the infinite variance regressors are consistent at a rate of $T^{1/\alpha}$,  still faster than the usual rate of $\sqrt{T}$.  We then show that the disaster shock series can be identified by the magnitude of its  kurtosis and  the sign of its impact effect. For estimation, we perform an independent components analysis (ICA) based on distance covariance of the  pre-whitened data, an approach first  suggested in \citet{matteson-tsay:17}  for  finite variance data.  \citet{davisfernandes2021} recently showed that the procedure  remains valid   when  a shock has infinite variance provided its mean is finite.

Prewhitening by singular value decomposition is often used  to  remove correlations prior to ICA estimation  to focus on the higher order signals.  For  SVAR applications, prewhitening by  Choleski decomposition  is more natural since it is already used to identify mutually uncorrelated shocks with a recursive structure. We show that even though the  variance of the  shocks  may not exist,   Choleski decomposition of  the sample covariance  remains valid. Furthermore, we show that  ICA  will still recover the shocks  in spite of sampling uncertainty in the VAR residuals. To assess the restrictions imposed on the SVAR,    we  apply a permutation-based procedure   to the distance covariance statistic as a test for independence that is    robust  to  infinite variance data. It complements  other SVAR specification tests  made possible by the independence assumption, as discussed below.

The rest of the paper is structured as follows.  Section 2 summarizes the key properties of heavy-tailed linear processes and discusses the implications for VAR estimation. Section 3 presents the HL  framework. Consistency and limiting behavior of the least squares estimator for parameters in a VAR are shown. Identification,  estimation  via   distance covariance, and  implementation of  an independence test  are then discussed.  Section 4 uses simulations and three applications to assess the properties of the proposed procedure.     The appendix contains background material on distance covariance as well as proofs of the main results in Section 3.



\section{Heavy-Tailed Linear Processes}
Disaster events are rare and heavy tails can be a useful  characterization of their probabilistic structure. Well known heavy-tailed distributions  include  the Student-$t$, $F$, Fr\'echet, as well as infinite variance stable and Pareto distributions. 

Let $F(x)=P(Z\le x)$ for $x\in\mathbb R$ be the distribution of an IID sequence of  random variables  $\{Z_t,t=0,\pm 1,\pm2, \ldots\}$. Then $F$ has Pareto-like tails with tail index $\alpha>0$ if 
\begin{equation}
\label{eq:bd13.3.1}  x^{\alpha}\mathbb P(|Z|>x)\rightarrow C, \quad   x\rightarrow\infty\,,
\end{equation}     
where $C$ is a finite and positive constant and $\frac{\mathbb P(Z>x)}{\mathbb P(|Z|>x)}\rightarrow  p\in[0,1]$ as $x\rightarrow\infty$. 
 Examples include the Cauchy  and Pareto distributions.  The Gaussian distribution has `thin' tails that decay faster than an exponential  and is not included in this class. The results that follow can be extended to a more general condition on $F$ called regular variation in which \eqref{eq:bd13.3.1} is replaced by
\begin{equation*}
 \frac{\mathbb P(|Z|>sx)}{\mathbb P(|Z|> s)}\rightarrow x^{-\alpha}, \quad   s\rightarrow\infty\,.
\end{equation*}
The normalizing constants in such an extension become less explicit so we stick to the Pareto-like tail assumption for tractability. 

Let $d_{1T}=\inf \{ x:\mathbb P(|Z_1|>x)\le \frac{1}{T}\}$ be the $(1-\frac{1}{T})$-th   quantile of $F$ and $d_{2T}=\text{inf}\{ x:\mathbb P(|Z_0Z_1|>x)\le T^{-1}\}$ be the corresponding quantile for the joint distribution of the product $Z_0Z_1$. Distributions with Pareto-like tails have $d_{1T}=T^{1/\alpha}C^{1/\alpha}$.
  Since $1-F(d_{1T})=1/T$ for continuous $F$,   (\ref{eq:bd13.3.1}) implies
\begin{equation}
\label{eq:eq3}
T \mathbb P(|Z_t|>d_{1T} x)\rightarrow x^{-\alpha},\quad \mbox{as $T\rightarrow\infty$\,,}
\end{equation}
for all $x>0$.  Similarly $T \mathbb P(|Z_0Z_1|>d_{2T} x)\rightarrow x^{-\alpha}$ (see {\citet{davis-resnick:86}}).  The population   moments of $Z_t$ satisfying  (\ref{eq:bd13.3.1})  are only defined   up to order $\alpha$  since
\[ \begin{cases} \mathbb E|Z_t|^\delta   = \infty \quad \delta \ge \alpha, \\
\mathbb E|Z_t|^\delta < \infty \quad \delta<\alpha.
\end{cases}
\]
 It is possible for  the population variance to exist but  the population kurtosis to be undefined. But  even if the population moments do not exist,  the sample moments can still have well defined limits. 
If $Z_t$ has a Pareto-like tails with index $\alpha\in (1,2)$, $Z_t^2$ also has Pareto-like tails with index $\alpha/2$, and it holds that
\begin{eqnarray} \begin{pmatrix*}[r] d_{1T}^{-1}\sum_{t=1}^T (Z_t-E Z), &  d_{1T}^{-2} \sum_{t=1}^T Z_t^2, & d_{2T}^{-1}\sum_{t=1}^T (Z_t-\bar Z) (Z_{t-h}-\bar Z)\end{pmatrix*} \dconv \begin{pmatrix} S_{\alpha}, & S_{\alpha/2,0}, & S_{\alpha,h}\end{pmatrix}
\label{eq:partialsums}
\end{eqnarray}
where $\bar Z=\frac1T\sum_{t=1}^TZ_t$ is the sample mean, and for $h> 0$, $S_{\alpha}, S_{\alpha/2,0},  S_{\alpha,h}$ are stable random variables with exponents, $\alpha,\alpha/2$, and $\alpha$ respectively.  Their joint distributions can be found in \citet{davis-resnick:86}.

To gain a sense of the tail properties of the data under investigation, we  will  make use of the fact that if
 $Z_t$   is an IID Pareto sequence with tail index $\alpha=1$, then the sample kurtosis $\kappa_4$ has the property (see \cite{cohen2020heavy}) that
\begin{equation}\label{eq:kurtosis}
\frac{1}{T}\kappa_4= \frac{ \sum_{t=1}^T Z_t^4}{(\sum_{t=1}^T Z_t^2)^2 }\dconv  \frac{S_{\alpha/4}}{S_{\alpha/2}^2}.
\end{equation}
The limit of kurtosis, scaled by the sample size, is a random variable between zero and one so the maximum kurtosis that can be observed asymptotically is  $T$. Tabulating the distribution for $T=500$ and $T=1000$ with $\alpha=1$, we see that the quantiles roughly double with $T$. Based on simulations, the values of these quantiles are an upper bound for $\alpha\in (1,2)$.

\begin{center}
\begin{tabular}{l|lllllllll}
$T$ &    1\% &      2.5\% &        5\%  &      10\%  &  50\% &   90\%  &      95\%   &    97.5\%     &  99\%  \\ \hline
500 & 37.9 & 47.55 &  58.8 & 77.9& 225.9 &  477.5 & 494/5 & 498.7 & 499.7 \\
1000 & 76.8 & 95.4 & 119.0 & 151.7 & 445.6 & 952.8 & 987.4 & 996.8 & 999.4\\ \hline
\end{tabular}
\end{center}
As a point of reference,  the disaster series shown in Figure \ref{fig:cddd} has $T=480$ and kurtosis of around 70, which is in  the lower 10-th percentile.\footnote{The distribution of $S_\alpha$ can be approximated by simulating $j=1,\ldots J$ times $s_{j,\alpha}=\sum_{m=1}^M (\sum_{j=1}^m e_j)^{-1/\alpha}$  where $\{e_j\}$ is drawn from the exponential distribution.}  The number of deaths series mentioned in the Introduction has kurtosis of 147 and is in the 30-th percentile. In contrast, the kurtosis a  typical of macro economic time series is under 10, hence the theory for heavy tails would be inappropriate. A multivariate system of time series with different tail properties thus necessitates a different setup.

There is a large literature on robust and quantile  estimation of the parameters in a linear model to guard against  extreme values which explicitly    down-weights  outliers.  \citet{blattberg-sargent:71} and \citet{kadiyala:72} show that the least squares estimator is unbiased when the error in the regression model is drawn from a  general symmetric stable Paretian distribution, but it is not  the  best linear unbiased estimator. In the Cauchy case when $\alpha=1$, the best linear unbiased estimator  is  $y_\tau/x_\tau$ where $x_\tau=\max_j X_j$.\footnote{Best here means in terms of minimizing dispersion.}  A different viewpoint, also the one taken in this paper,  is that the  extreme values are of interest.\footnote{See, for example, two special issues on heavy-tailed data, \citet{heavytails-joe:13} and \citet{heavytails-joe:14}.}   Under this assumption and  fixed regressors, \citet{mikosch-devries:13} provide a finite sample analysis of the tail probabilities of  the single equation CAPM estimates to understand why they vary significantly across reported studies.  We are interested in estimating  dynamic causal effects   in a multivariate   setting when  the  regressors are stochastic, and one of the primitive shocks has   heavy tails.

\subsection{Implications for VAR Estimation}
Consider $n$  mean zero variables $Y_t=(Y_{t1},\ldots, Y_{tn})^\prime$ represented by  a   VAR(p):
\begin{eqnarray*}
 Y_t&=& A_1 Y_{t-1}+\ldots+A_p Y_{t-p}+e_t, 
\end{eqnarray*}
where $A(z) =I_n-A_1 z -\ldots -A_p z^p$ is the matrix-valued AR polynomial.    Provided that det$A(z)\ne 0$ for all $z\in \mathbb C$ such that  $|z|\le 1$, $A(z)^{-1}$ exists, the   moving-average representation of the model is  $ Y_t= \Phi(L) e_t $ where  $L$ is the lag operator, and $\Phi(L)=A(L)^{-1}$ with $\Phi_0=I_n$.    

The standard OLS estimator $\hat A$ of $A$ is  characterized by  (see \eqref{eq:ahatols} and \eqref{eq:ahatm}) 
\begin{eqnarray*}
\hat A-A &=&\bigg(\sum_{t=1}^T e_{t} Y_{t-1}^\prime\bigg) \bigg(\sum_{t=1}^T Y_{t-1} Y_{t-1}^\prime\bigg)^{-1}.
\end{eqnarray*}
 These errors $e_t$ are mapped to a $n\times 1$ vector of  primitive shocks $u_t=(u_{t1},\ldots, u_{tn})^\prime$ via 
 a (time invariant)  matrix $B$:
\begin{equation*}
e_t= B u_t\,,
\end{equation*}
where $u_t$  is usually assumed to be mean zero, mutually and serially uncorrelated and with  $\Sigma_u=\mathbb E[u_tu_t^\prime]$ being a diagonal matrix. See, for example, \citet{stock-watson:macrohandbook2} and \citet{kilian-lutkepohl:book}. The reduced form errors $e_t$ are usually assumed to have `light tails' which is possible  only if  $u_t$ has light tails. A model that satisfies these standard assumptions  will be referred to as the LL (light-light)  hereafter. Under regularity conditions for least squares estimation,   $\hat A$ is $\sqrt{T}$ consistent and asymptotically normal. 

 The modeling issues that arise when one of the primitive shocks in a SVAR has infinite variance are best understood in the   $p=1$ and $n=2$ case.  Consider first a  HH (heavy-heavy) model in which  both shocks have heavy tails. 

\begin{lemma} 
\label{lem:lemma1}
 Let $\{Z_t\}$ be an IID  sequence of random variables with Pareto-like tails (i.e., equation \eqref{eq:bd13.3.1}) with index $\alpha\in (0,2)$ and $EZ_t=0$ if $\alpha>1$. Let 
\[d_{1T}=T^{1/\alpha}, \quad\quad d_{2T}=(T\log T)^{1/\alpha}.
\]
 If the sequence of constants $\{\psi_j\}$ are such that $\sum_{j=-\infty}^\infty |\psi_j|^\delta<\infty$  for some $\delta\in(0,\alpha)\cup[0,1]$, then 
\begin{itemize}
\item[i.] the process  $X_t=\sum_{j=-\infty}^\infty \psi_j Z_{t-j}$ exists with probability one and is strictly stationary. 
\item[ii.] Let $\hat\rho(h)=\sum_{t=1}^{T-h} X_tX_{t-h}/\sum_{t=1}^T X_t^2$ be the sample  autocorrelation  at lag $h>0$ and suppose that   $\sum_{j=-\infty}^\infty |j|\; |\psi_j|^\delta<\infty$ for some $\delta\in(0,\alpha)\cup[0,1]$. Then for $\alpha\ne 1$,
\begin{eqnarray*}
(d_{1T}^{-2} \sum_{t=1}^T X_t^2,\; d_{2T}^{-1} \sum_{t=1}^{T-h} X_tX_{t-h} )
&\dconv& (S_{\alpha/2,0},S_{\alpha,h})\\
\bigg(\frac{T}{\log T}\bigg)^{1/\alpha}( \hat \rho(h)-\rho(h))&=& \frac{S_{\alpha,h}}{S_{\alpha/2,0}}
\end{eqnarray*}

where   $(S_{\alpha/2,0}$, $S_{\alpha,h}$) are  independent  stable random variables with indices $\alpha/2$ and $\alpha$, respectively.
If $\alpha>1$, then the latter convergence also holds if $EZ_t\ne 0$ provided $\hat\rho(h)$ is replaced by its mean-corrected version, $\tilde\rho(h)=\sum_{t=1}^{T-h}( X_t-{\bar X})(X_{t-h}-{\bar X})/\sum_{t=1}^T (X_t-{\bar X})^2$.

\item[iii.] $d_{2T}^{-1}\sum_{t=1}^T X_{t-1} Z_t\dconv S_{\alpha}$, where $S_{\alpha}$ is a  stable random variable with index $\alpha$. 
\end{itemize}
\end{lemma}
By restricting attention to $0<\alpha<2$, we only consider processes with infinite variance.
  Even though $X_t$ is not covariance stationary (since $\mathbb E|X_t|^2=\infty$), part (i) states that the process $X_t$ exists and is strictly stationary. 
  The stated results for the sample covariance and sample autocorrelation  are due to  \citet[Theorem 3.3]{davis-resnick:86}  and also hold  when $X_t$ is centered for $\alpha\in (1,2)$. Note that the convergence  of $\hat \rho(h)$ is faster than the $\sqrt{T}$ rate obtained for finite innovation variance.

For VAR estimation,  Lemma \ref{lem:lemma1} can be used to show that
\begin{eqnarray*}
d_{2T}^{-1}  \sum_{t=1}^T B u_tY_{t-1}' &=&\sum_{h=0}^\infty B \bigg( (T\log T)^{-1/\alpha} \sum_{t=1}^T  u_t  u_{t-1-h}\bigg) \Psi_h^\prime 
\dconv  \sum_{h=0}^\infty B S_{uu,h}\Psi_h^\prime =: S_{Ye} \\  
d_{1T}^{-2}  \sum_{t=1}^T Y_{t-1} Y_{t-1}^\prime &=& \sum_{h=0}^\infty \Psi_h \bigg(T^{-2/\alpha}\sum_{t=1}^T u_tu_t^\prime\bigg) \Psi_h^\prime
 \dconv \sum_{h=0}^\infty \Psi_h   S_{uu,0}\Psi_h^\prime=: S_{YY}.
\end{eqnarray*}
It  then follows from continuous mapping  that  the least squares estimator  is super consistent:
\begin{eqnarray*}
\bigg(\frac{T}{\log T}\bigg)^{1/\alpha}(\hat A-A)&\dconv& S_{Ye} S_{YY}^{-1} .
\end{eqnarray*}
Though the analysis is straightforward, this setup  is  unappealing for macroeconomic data  because if $u_{t1}$ and $u_{t2}$ both have infinite variance,   $Y_{t1}$ and $Y_{t2}$ must also  have infinite variance. But  a  typical economic time series does not resemble the  series shown in Figure \ref{fig:cddd}.  Not only is the disaster series  much less persistent, its kurtosis (over 70) is an order of magnitude larger than for variables like output growth, inflation, and interest rates. 

\section{A VAR with Heavy and Light  Tailed Shocks}

 Our goal is  a model in which (i) a heavy tailed shock $u_{t1}$ co-exists with  light tailed shocks $u_{ti},\,i=2,\ldots,n$, and (ii)  $Y_{ti}$ is influenced by the current and past values of $u_{t1}$  but not dominated by them in a sense to be made precise.  We consider the HL (heavy-light)   model derived from the SVAR$(p)$ 
\begin{eqnarray}\label{eq:var}
	Y_t&=& A_1 Y_{t-1}+\ldots+A_p Y_{t-p}+Bu_t, 
\end{eqnarray}
where for each $h=1,\ldots, p$, $A_h$ is a $n\times n$ matrix with $(i,j)$-th entry denoted $[A_{ij}^{(h)}]$,  the coefficient of variable $j$ at lag $h$ in equation $i$. The entries $[B_{ij}]$ of the $n\times n$ matrix $B$ are similarly defined.
\paragraph{Assumption HL}  
\begin{itemize}

\item[i.] The  sequence of $n$-dimension random vectors $\{u_t\}$ is iid and the components, $u_{ti}\,, i=1,\ldots,n$ are also independent. The $u_{t1}$ will have Pareto-like tails with  index $1<\alpha<2$ and  $\mathbb E[u_{t1}]=0$, while the remaining shocks $u_{ti}\,,i=2\ldots,n$ will have  thin tails with mean zero and variance 1.  
 
\item[ii] The coefficient matrices $A_h$ for $h=1,\ldots, p$ and the matrix $B$ will satisfy the following conditions.
\begin{eqnarray}
A_{i1, T}^{(h)}=a_{i1}^{(h)}/T^{\theta}\,,~~i=2,\ldots, n,\label{eq:A}\\
B_{i1, T}=b_{i1}/T^{\theta}\,,~~i=2,\ldots, n,\label{eq:B}
\end{eqnarray}
 with 
\begin{equation}\label{eq:theta}
\theta=\frac{1}{\alpha}-\frac{1}{2}.
\end{equation} 
\end{itemize}

The primitive shocks   $u_{ti}$  are assumed to be independent  across $i$ and $t$ but  does not preclude  time varying second moments, though it is stronger than mutual orthogonality of $u_{t}$ typically  assumed in SVAR modeling.
Assumption HL (i) restricts attention to processes with  tail index $1<\alpha<2$ and thus excludes Cauchy shocks.  The assumption that the thin tailed shocks $u_{ti}\,,i=2,\ldots n$ have  unit variance is without loss of generality, but it is important that their variances are finite.  Since the variations of $u_{t1}$ will dominate those of $u_{ti}\,,i\ge2$  when both are present,  $Y_{t1}$ will have heavy tails and exhibit the large spikes originating from $u_{t1}$. 

Assumption HL(ii) is motivated by the fact that $Y_{ti}$ cannot have finite variance  unless   $B_{i1}=0$ and $A_{i1}^{(h)}=0$ for all $h$. But the dynamic effects of $u_{t1}$ on $Y_{t+h,i}$ would then be zero at all lags by assumption, rendering the empirical exercise meaningless. Thus,
  the $Y_{ti}$ equation is modified to dampen the influence of $u_{t1}$ on $Y_{ti}$ at  rate $\theta$ given in \eqref{eq:theta}, so that  $T^{-1/2} \sum_{t=1}^T u_{t1}/T^\theta=T^{-1/\alpha} \sum_{t=1}^T u_{t1}$ has a limit.\footnote{A similar effect can be achieved by replacing the heavy-tailed shock by its truncated version $\frac{u_{t1}}{T^\theta}1_{|u_{t1}|\le MT^{1/\alpha}}$ for some constant $M$.  This was the approach taken by \citet{amsler-schmidt:12}.}  Localizing $A_{i1,T}^{(h)}$ and  $B_{i1,T}$ to zero is an asymptotic device  to obtain this limit, but note that $A^h_{i1,T}$ and $B_{i1,T}$ are not time varying. Under assumption HL(ii), $Y_{t,T}$ is  a triangular array that depends on $T$. To simplify notation,  the explicit dependence on $T$ is suppressed. 

 A heavy-tailed linear time series must have a heavy-tailed shock as its  primary source of variation, but it need not be exogenous. In our model, exogeneity would require that $A_{1j}=B_{1j}=0,\,j=2,\ldots,n$,  in which case, any feedback from $Y_{ti}$,  $i\ge2$ to $Y_{t1}$ would be disabled. But such a  model would  not shed light on  how  macroeconomic outcomes might  mitigate or amplify the  effects of disasters. Assumption HL allows $A^{(h)}_{1j}$ and $B_{1j},\,j=2,\ldots n$ to be free parameters to be estimated.

Specializing to the $n=2$ and $p=1$ case with $Eu_{t1}^2=\infty$ and $Eu_{t2}^2<\infty$ 
we show in the Appendix that the following holds under Assumption  HL 
\begin{eqnarray*}
 \frac{1}{\sqrt{T}} \sum_{t=1}^T Y_{t2}&=&\dconv  S^*_{Y,2}\\
\frac{1}{T} \sum_{t=1}^T Y_{t2}^2&=&  \dconv S^*_{YY,22},
\end{eqnarray*}
where the limits have a stable distribution with index $\alpha$ and $\alpha/2$, respectively.  Thus  the sample first and second moments of $Y_{t2}$ have (random and possibly constant) limits    even though  one of its shocks has infinite variance. The implications for least squares estimation of the HL model can be summarized as follows.

 \begin{proposition} 
\label{prop:prop1}
Suppose that the data are generated by  \eqref{eq:var}, and for tractability assume that  $n=2, p=1$ with VAR(1) coefficient matrix $A=[A_{ij}]_{i,j=1}$. If Assumption HL holds,  then the least squares estimate of $A$ satisfies
\begin{eqnarray*}
 \sqrt{T} (\hat A_{11}-A_{11}) &\dconv& S_{A,11}\\
T^{1-1/\alpha}(\hat A_{12}-A_{12})&\dconv& S_{A,12}\\
T^{1/\alpha}(\hat A_{21}-A_{21})  &\dconv & S_{A,21}\\
 \sqrt{T} (\hat A_{22}-A_{22}) &\dconv & S_{A,22}.
\end{eqnarray*}
\end{proposition}
The convergence rate for $\hat A_{11}$  is  $\min(\frac{T}{\log T}^{1/\alpha},T^{1/2})$,  which is $\sqrt{T}$. This  is slower than  the rate for $\hat A_{11}$ in  the HH model  because  one of the infinite variance regressors in the HH model is  replaced by one that has  finite variance.  The convergence rate for  $\hat A_{12}$  can be written as $\sqrt{T} T^{-\theta}$ which is slower than the $\sqrt{T}$ rate for $\hat A_{12}$ in the LL model  because the variations in  this equation are dominated by those from lags of $Y_{t1}$, hampering identification of $A_{12}$. Now the convergence rate for $\hat A_{21}$ can be written as  $\sqrt{T}T^\theta$ which is faster than the $\sqrt{T}$ rate obtained for $\hat A_{21}$ in the LL model. This implies that $\hat a_{21}=T^\theta \hat A_{21}$  is $\sqrt{T}$ consistent.  Hence in the HL model, the local parameter $a_{21}$ is consistently estimable.  In each case, the limit distribution is non-standard and not pivotal, so that construction of asymptotically correct confidence intervals is intractable.

Since VAR  estimates are obtained from  least squares regressions on an equation by equation basis,  Proposition \ref{prop:prop1} sheds light on the more general setting when  a regressor has infinite variance, but the dependent variable has finite variance. Though such a regression would  be `imbalanced' in the standard setup,     the coefficients on the heavy-tailed variable  is being scaled down to accommodate the heavy-tailed shock in our HL setup. The coefficient  estimate on the infinite variance regressor would be consistent but not asymptotically normal. By implication, the impulse response coefficients whether computed from the VAR or by local projections would likely not be asymptotically normal.

\subsection{Identification of $B$ and the Primitive Shocks}\label{identification}
The structural moving-average representation of the model is
\begin{eqnarray*}
Y_t&=&\Phi(L) B u_t=\Psi(L) u_t\\
&=&\Psi_0u_t + \Psi_1 u_{t-1}+\ldots,
\end{eqnarray*}
where  $\Psi(L)=A(L)^{-1}B$ and  $\Psi_0=B$.  The  effects of $u_{t1}$ on $Y_{t+h,2}$ are given by the first column of $\Psi_h$ which depends on $A$ and $B$. Hence to estimate the dynamic causal effects of $u_{t1}$,  we  need  to be able to consistently estimate $B$ when $u_{1t}$ has infinite variance. 

 The relationship between the vector of primitive shocks $u$ and error terms is
\begin{equation}\label{eq:ica}
e_t=Bu_t  ~~~\mbox{and}~~~ u_t=We_t 
\end{equation}
where $u_t=(u_{t1},\ldots,u_{tn})'$ is an $n$-vector consisting of independent random variables with mean zero and $B$ is an $n\times n$ matrix with inverse $W$. 
As is well known,   $B$  is not uniquely identified from the second moments of $e_t$ alone  even when  $e_t$ has finite variance  because $BQ'Qu_t$ has the same covariance structure as $Bu_t$ for any orthonormal matrix $Q$. 


\begin{lemma}
	\label{lem:IDtest}
	Let $e=Bu$, where $u$ is  a $n\times 1$ vector of mutually independent components of which at most one is Gaussian and $B$ is an $n\times n$ invertible matrix with inverse $W=B^{-1}$.  If the components of $\hat u=\hat W  e$ are pairwise independent, where $\hat W$ is an invertible matrix, then $\hat W = P\Lambda W$ where  $P$ is a permutation matrix and $\Lambda$ is a diagonal matrix.  Further, the components of $\hat u$ must be mutually independent. 
\end{lemma}

\medskip
\noindent{\bf Proof.}  The proof of this result follows directly from Skitovich-Darmois theorem as described in the proof of Theorem 10 in  \citet{comon:94}.   Since  $\hat u=\hat W B u=:Gu$, the components of $\hat u_i$ can be written as
$$
\hat u_i=\sum_{k=1}^n G_{i,k}u_k.
$$
The independence of the $\hat u_i$ and $\hat u_j$ components implies that $G_{i,1}G_{j,1}=0$ for $i\ne j$. That is, the first column of $G$ contains at most one nonzero value.  A similar conclusion holds for all the columns of $G$.  Hence $G$ is product of a permutation matrix $P$  times a diagonal matrix $\Lambda=\text{diag}\{\lambda_1,\ldots,\lambda_n\}$, i.e., $G=P\Lambda$.  In other words, $\hat W W^{-1}=P\Lambda$ or $\hat W=P\Lambda W$ as was to be shown.  It follows by the form of $G$ that the components of $\hat u$ must be mutually independent.  $\Box$

%

Independence of $u$ narrows the class of  observational equivalent models to those characterized by permutations of rows and changes of scale/sign.   As discussed in \citet{gmr:17},   scale changes are responsible for failure of local identification, a problem that can be dealt with by normalizing the shocks   so that $\Lambda$ is an identity matrix. Failure of global identification arising from  permutation and sign changes require additional assumptions. It is only when the restrictions are correctly imposed that $P$ is also an identity matrix, in which case,    $\hat W=W$.

We also need to  impose restrictions on $W$ to identify a component of $u$ as a disaster shock.     Our problem is non-standard because the shock of interest   has a  heavy tail, but this distinctive  feature  actually helps identification. We reorder the components by their tail-heaviness, and take the disaster shock to be the first component, which is also the one with the largest kurtosis.  In practice, the variables in the estimated $u$ will be ordered by sample kurtosis. As seen in \eqref{eq:kurtosis}, this ordering is consistent with ordering the components of $\hat u$ by tail-heaviness.



\subsection{ICA and Distance and Covariance Estimation of $u$}\label{sec:4}


  Independent  components analysis (ICA) is  widely used  to identify a linear mixture of non-Gaussian signals. Whereas PCA uses the sample covariance to find uncorrelated signals, ICA   typically uses properties of the random vector that go beyond second moment properties in order  to separate the independent signals.\footnote{The two will give similar results when the higher-order statistics add little information. For a recent review, see   \citet{hyvarinen:13}.}  In the ICA literature, $B$  is  known as the {\it mixing} matrix and $W$   the  {\it unmixing} matrix. 
 ICA has been applied to  finite variance SVARs  in which  global identification is achieved by imposing additional  restrictions such as  lower triangularity of $B$.\footnote{ See, for example,  \citet{moneta-el:13, hzsh:10, gmr:17, maxand:19}, \citet{lanne-meitz-saikkonen:17}.}   




There exist many ICA estimators for identifying the source process, which in our case corresponds to   the primitive shocks $u$. Some procedures evaluate negative entropy (also known as negentropy) and take as solution the $W$ that maximizes non-Gaussianity  of $We_t$, while others   maximize an approximate likelihood using, for example, log-concave densities.  The popular fast ICA algorithm of \citet{hyvarinen-book} is a fixed-point algorithm for pseudo  maximum-likelihood estimation.  A different class of procedures  take as the starting point that if the signals are  mutually independent at any given $t$, their  joint density, if it exists,  factorizes into the product of their marginals. This suggests to evaluate the distance between the joint density  and the product of the marginals.\footnote{See, for example, \citet{bach-jordan:01} and \citet{eriksson-koivunen:03},  and
\citet{hyvaninen-oja:00} for an overview of the methods used in signal processing. Statistical procedures include \citet{chen-bickel:06}, \citet{hastie-tibshirani:03}, \citet{hastie-tibshirani:03}, \citet{samworth-yuan:12}, \citet{gmr:17}.}  {\citet{chen-bickel:06} form a distance measure between the joint characteristic function and the product of the marginal characteristic functions to estimate the unmixing matrix.  The advantage of this procedure is that it does rely on existence of joint densities or moments.  In case the vector has finite second moments, they obtain a convergence rate of  $1/\sqrt{T}$ for this nonparametric estimate of $W$, the same as the one obtained in \citet{gmr:17} for  parametric estimation.   \citet{matteson-tsay:17}  use a {\em distance covariance} approach to extract the independent sources under the  assumption that they  have finite variances, which is similar in spirit to the method of \citet{chen-bickel:06}. 

\begin{remark}\label{rmk:Omega} 
Following \citet{chen-bickel:06} we assume that the parameter space of unmixing matrices is given by $\Omega$ consisting of invertible matrices $W$ for which a) each of its rows has norm 1; b) the element with maximal modulus in each row is positive; c) the rows are ordered by $\prec$; for  $a,b\in \bbr^n$, $a\prec b$ if and only if there exits $k\in \{1,\ldots,n\}$ such that $a_i=b_i$, $i=1,\ldots,k-1$ and $a_k<b_k$.  Further it is assumed that the true unmixing matrix $W_0\in \Omega$. However, we will reorder the rows of the estimated $W$ according to largest sample kurtosis.  The disaster shock with infinite variance will correspond to the first row of $W$. 
\end{remark}

We will also use the distance covariance approach because as shown in the companion paper \citet{davisfernandes2021}, it is also valid when one component of $u$ has infinite variance.   The distance covariance between two random vectors $X$ and  $Y$ of dimensions $m$ and $n$, respectively, is 
\begin{equation}\label{eq:distcov}
{\mathcal I}(X,Y;w)= \int_{\bbr^{m+n}} \big|\varphi_{X,Y}(s,t)-\varphi_X(s)\,\varphi_Y(t)\big|^2 w(s,t)ds\,dt\,,
\end{equation}
where $w(s,t)>0$ is a weight function and $\varphi_Z(t) = E [\exp^{i (t,Z)}],\, t\in\bbr^d \,$ denotes the characteristic function for any random vector
$Z\in\bbr^d$.  The most commonly used weight function, which we will also adopt here, is
\begin{equation}\label{eq:szekely}
w(s,t)=\left(c_{m,\beta}|s|^{\beta+m}c_{n,\beta}|t|^{\beta+n}\right)^{-1}\,,
\end{equation}
where $\beta\in (0,2)$,  $c_{m,\beta}=\frac{2\pi{m/2}\Gamma(1-\beta/2)}{\beta2^\beta\Gamma((\beta+m)/2)}$  (see \citet{szekely-rizzo-bakirov:07}). The integral in \eqref{eq:distcov} is then finite provided $E|X|^\beta +E|Y|^\beta<\infty$.  Under this moment assumption, one sees immediately that $X$ and $Y$ are independent if and only if ${\mathcal I}(X,Y;w)=0$ since in this case the joint characteristic function factors into the product of the respective marginal characteristic functions, $\varphi_{X,Y}(s,t)=\varphi_X(s)\,\varphi_Y(t)$ for all $(s,t)\in \bbr^{m+n}$.     Based on data $(X_1,Y_1)\,\ldots,(X_T,Y_T)$ from $(X,Y)$, the general distance covariance in \eqref{eq:distcov} can be estimated by replacing the characteristic functions with their empirical counterparts $\hat\varphi_{X,Y}, \hat\varphi_{X}$ and $\hat\varphi_{X,Y}$,  where e.g., $\hat\varphi_{X,Y}(s,t)=\frac{1}{T} \sum_{j=1}^T \exp\{i(s,X_j)+i(t,Y_j)\}$. Then 
\begin{eqnarray}
\hat{\mathcal I}(X,Y;w)= \int_{\bbr^{m+n}} \big|\hat\varphi_{X,Y}(s,t)-\hat\varphi_X(s)\,\hat\varphi_Y(t)\big|^2 w(s,t)ds\,dt
\end{eqnarray}
Using the  $w$  given in \eqref{eq:szekely} and assuming $E|X|^\beta|Y|^\beta<\infty$, there is an explicit formula for $\hat{\mathcal I}$ (see \eqref{eq:distcovhat}) that avoids direct computation of the associated integral.   
Additional background on distance covariance can be found in the Appendix.


 Now the components of say a random vector  $S=(S_1,\ldots,S_n)'$ are independent if and only if ${\mathcal I}(S_k,S_{k+1:n},w) =0$ for $k=1,\ldots, n-1,$ where $S_{k+1:n}=(S_{k+1},\ldots,S_n)'$.   \citet{matteson-tsay:17} observe that the independence condition is equivalent to ${\mathcal I}_{MT}=0$, where
\begin{equation}\label{eq:matteson}
{\mathcal I}_{MT}= {\mathcal I}(S_{1},S_{2:n})+ {\mathcal I}(S_{2},S_{3:n})+\cdots+{\mathcal I}(S_{n-1},S_{n-1:n})\,,
\end{equation} 
with weight function given by \eqref{eq:szekely}. 
Based on a sample $e_t=(e_{t1},\ldots,e_{tn})',\,t=1,\ldots,T$, an estimate of the unmixing matrix $W$ is found by minimizing the objective function, 
\begin{equation}\label{eq:ica-obj}
\hat{\mathcal I}_{MT}(W):=\hat {\mathcal I}(S_{1},S_{2:n})+ \hat{\mathcal I}(S_{2},S_{3:n})+\cdots+\hat{\mathcal I}(S_{n-1},S_{n-1:n})\,,
\end{equation}
subject to $W\in\Omega$ and  where $\hat{\mathcal I}$ is the empirical estimate of $\mathcal I$ using  $S_t=We_t,~t=1,\ldots,T$. \citet{matteson-tsay:17}  show that  procedure  produces a consistent estimate of $W$ when the variance of the $S_t$ is finite. The proof is based on rewriting $\mathcal I(\cdot)$ in terms of $V$ statistics and presumes that terms of the form $E|XY|$ are finite. 

 In our case of infinite variance,  $\mathcal I(X,Y)$ is finite even if $E|XY|=\infty$.  One only needs that $E|X|+E|Y|<\infty$.  More recently, it is shown in \citet{davisfernandes2021} that consistency of $\hat W$ based on the sample distance covariance also holds in the infinite variance case. This result justifies the use of the objective function $\hat{\mathcal I}_{MT}(W)$ for estimating the unmixing matrix in the finite mean but infinite variance case. In case the mean is infinite, one can choose a $\beta<1$ in the weight function to ensure that the moment condition $E|e_t|^\beta<\infty$ is met. 

\subsection{Prewhitening and Choleski Decomposition}
In most ICA estimation procedures, the first step is typically to prewhiten the output. In effect, prewhitening removes second moment correlations prior to estimating the independent components. In the context of a SVAR with finite variance, suppose we have the observations  $e_1,\ldots,e_T$, from the model $e_t=Bu_t$.  Denote the sample covariance matrix of the $e_t$'s by $\hat \Sigma_e^{-1/2}$, from which it's prewhitened values are given by $\tilde e_t=\hat \Sigma_e^{-1/2}e_t$, where the inverse square root matrix is, for example, computed from the  singular value decomposition (SVD) of the sample covariance matrix.   Then one can restrict candidate unmixing matrices $W$ to have the form $W=O\hat \Sigma_e^{-1/2}$, where $O$ is an orthogonal matrix.  In particular, one could optimize the function in \eqref{eq:matteson}, e.g.,
\begin{equation}\label{eq:Ohat}
	\hat O=\mbox{argmin}_{O\in\cO{n}} \hat I_{MT}(O\hat\Sigma_e^{-1/2})\,,
\end{equation}	
where the minimization is over all $O\in \cO{n}$, the space of $n$-dimensional orthogonal matrices.  This produces an estimate of the unmixing matrix given by $\hat O\hat\Sigma_e^{-1/2}$ that is a consistent estimate of $W_0$ after  suitable rescaling and row permutation as noted in Remark \ref{rmk:Omega}. The optimization over orthogonal matrices reduces the number of unknowns from $n^2$ to $n(n-1)/2$.  The fact that this prewhitening step actually works in the infinite variance case follows directly from Theorem 3.2 in \citet{davisfernandes2021} (see also \citet{chen-bickel:06}), which we record in the following proposition.

\begin{proposition}\label{prop:propW}
Consider observations $e_t,\, t=1,\ldots, T,$ from the ICA model \eqref{eq:ica} where $W=W_0\in \Omega$ is the true unmixing matrix, and that the components of $u_t$ are mutually independent,  at most one has infinite variance, at most one component is normal and  none of the components are degenerate.
Then, setting $\hat W=[\hat O\hat \Sigma_e^{-1/2}]_\Omega$, the rescaled and row permuted version of  $\hat O\hat \Sigma_e^{-1/2}\in \Omega$, we have
\begin{equation*}
\hat W \cip W_0\,.
\end{equation*}
\end{proposition}

Although we have used the SVD version of $\hat \Sigma_e^{-1/2}$ in Proposition \ref{prop:propW}, we could also use the Choleski analogue, which is often  an attractive alternative.  This is especially true for SVARs since it is already widely used to identify a lower triangular structure of $B$.  Though the population covariance matrix of $e_t$ does not exist in the infinite variance case, a decomposition of  the sample covariance matrix is possible. The following result gives the decomposition for  the $n=2$ case.
 
\begin{lemma}
\label{lem:choleski}
Let $e^c_t=(e^c_{t1},e^c_{t2})'=\Sigma_e^{-1/2} e_t$, where $\Sigma_e^{-1/2}$ is the  Choleski decomposition of the sample covariance matrix $\hat\Sigma_e$, i.e., $\hat\Sigma_e=\Sigma^{1/2}(\Sigma^{1/2})^\prime$.
Under Assumption HL,    
\begin{eqnarray*}
	e^c_{t1} &\sim & T^{1/2-1/\alpha} |B^{-1}_{11}|S_{11,T}^{-1/2} e_{t1}\,,\\
	e^c_{t2} &\sim& |B_{22}|^{-1}\sigma_2^{-1} (e_{t2}-c_T  e_{t1})\,,
\end{eqnarray*} 
where $ c_T=\frac{\sum_{s=1}^Te_{s1}e_{s2}}{\sum_{s=1}^Te_{t2}^2}$, $S_{11,T}=T^{-2/\alpha}\sum_{t=1}^Tu_{t1}^2=O_p(1)$, and $\sigma_2^2=\mbox{var}(u_{t2})$.  Moreover,  $c_T\cip 0$ and hence $e^c_{t2}\approx sign(B_{22}) u_{t2}/\sigma_2$. 
\end{lemma}
The proof of the  lemma is given Section \ref{app:prooflem3}  of the Appendix. The prewhitened variables $e^c_t$ remain a function of $u_{t1}$ and $u_{t2}$ which we seek to identify. Observe that if $B$ were lower triangular, $e^c_{t1}$ will only depend on $u_{t1}$ since $e_{t1}=B_{11} u_{t1}+B_{12}u_{t2}$. But  note that  Choleski decomposition is used here only as a prewhitening device and not as a way to achieve  identification. If the ordering is incorrect, ICA  will undo the ordering to find the  $u$ satisfying the additional identification restrictions.


In practice, of course, we do not observe the  residuals $e_t$ directly but rather the estimated versions $\hat e_t=(\hat e_{t1},\ldots,\hat e_{tn})^T, \,t=1,\ldots,T$.  Limit distributions of the distance covariance function based on the residuals can be slightly different when applied to the $\hat e_t$ than the actual residuals (see \citet{davis2018applications}).  Interestingly, in the heavy-tailed case, the limit theory for the distance covariance based on estimated and actual residuals is the same.  In the context of consistency in the estimation of the unmixing matrix, the same procedure can be  carried out as above using estimated residuals  $\hat e_t=e_t+(A-\hat A)Y_{t-1}$.  

\begin{proposition}\label{prop:propWehat}
		Consider the estimated residuals $\hat e_t,\, t=1,\ldots, T,$ based on fitting the AR coefficients in an SVAR model using OLS.  Assuming the same  model framework as in Propositions \ref{prop:prop1} and \ref{prop:propW}, set $\hat W=[\hat O\hat \Sigma_{\hat e}^{-1/2}]_\Omega$, where the objective function in \eqref{eq:Ohat} is based on the estimated residuals instead of $e_t$.  Then
		\begin{equation*}
		\hat W \cip W_0\,.
		\end{equation*}
	\end{proposition}

The proof is given in the appendix.  The idea is that the sample residuals  can be represented by an ICA model with noise, i.e., $\hat e_t=Bu_t+v_t$    where the  noise is the sampling error $v_t=(A-\hat A(T))Y_{t-1}$.
It is then shown that the difference between $\hat \Sigma_e^{-1}$ (from noiseless model) and $\hat \Sigma^{-1}_{\hat e}$ (from noisy model) converges to zero in probability and thus  has asymptotically negligible effects on the objective function that estimates $W$. Applying  Theorem 3.3 in \citet{davisfernandes2021} for ICA with noise gives the stated result.




\subsection{An Independence Test of SVAR Restrictions}\label{sec:independence}


The dynamic properties of a SVAR are determined by restrictions imposed on the model and generally difficult to test. But if   $u_t$ is  independent, then by Lemma \ref{lem:IDtest},  the identifying restrictions are testable.  \citet{lanne-meitz-saikkonen:17} suggest  a procedure to first verify identifiability of the model, and  then   parametrically specify  $u$ (with finite variance) so that after maximum likelihood estimation,  the restrictions can be validated using classical  Wald and Likelihood ratio tests. QQ-plots of the identified shocks provide an additional check for  non-normality. \citet{herwatz:19} applies  a non-parametric test for independence  to ascertain whether  demand shocks  have no long run effects using bootstrap critical values.    \citet{amengual-fiorentini-sentana} use the  influence functions of a discrete mixture-normal  likelihood to test  the second, third, and fourth cross-moments while  explicitly accounting for  sampling uncertainty in $\hat e$. 
 

We also test independence of $\hat u$, but independence of  $\hat e$  is also  of interest because   if the components of $e_t=Bu_t$ were already independent, then by Lemma \ref{lem:IDtest},  $B$ would be diagonal and no  further  analysis on the structure of $W$ would be required.  A independence test of $\hat e$  is thus informative about its unrestricted  structure. In contrast, independence of $\hat u$ is informative about the structure implied by identifying restrictions. If $\hat u_t$ should fail an independence test, there would  be no point in further analyzing the impulse responses. 

As reviewed in \citet{josse-holmes:16},   many independence tests are available,  and if one suspects that the data have  features such as heteroskedasticity that are inconsistent with independence,  tests that target those features should have more power,   as in \citet{mopm:22}.  But we need a test that is also consistent when one component of $u_t$ has heavy tails. That is, the  test should  reject    with probability tending to one  as $T\rightarrow\infty$ for any $\hat W\ne W $ modulo permutations and scale/sign changes, irrespective of the tail properties of the data to be tested.  A test using  the empirical version of the aggregated distance covariance ${\mathcal I}_{MT}$ defined in \eqref{eq:matteson} can in principle be used. Even though $T\hat{\mathcal I}_{MT}$ has a limit distribution under the null hypothesis of independent components, the limit distribution is generally intractable.  Hence direct use of the limit distribution for  calculating cutoff values for the test statistic is infeasible.  

However, as pointed out in  \citet{matteson-tsay:17},
 one can use a test by  calculating the test statistic $\hat {\mathcal I}_{MT}$ for random permutations of the data. A permutation-based  test for independence is founded on the idea that if there is dependence in the components, then the value of $\hat{\mathcal I}_{MT}$  should be larger than the corresponding statistics based on  random permutations of the components, in which the dependence among the components has been severed by the permutation. The  test  is known to control Type I error and also  robust to the possibility of  heavy tails. Precisely, if $S_1,\ldots,S_T$ is an iid sample of random vectors of dimension $n$, then the permutation procedure is implemented via the following steps. For $b=1,\ldots NP$, 
\begin{itemize}
	\item[a.] For $j\in[1,n]$, generate   $(S^{(b)}_{t,j},t=1,\ldots T)=(S_{\tau_t^{(j)},j},t=1,\ldots T)$, where $\tau^{(j)}$ is a random permutation of $\{1,\ldots,T\}$.
	\item[b.] Compute $\hat{\mathcal I}^{(b)}_{MT}$ using $S^{(b)}$.
\end{itemize}
The test is distribution free under the null hypothesis.
The   $p$ value of the test is constructed as
\[ p(NP)= \frac{(k+1)}{(NP+1)}\]  where $k$ is the number of $\hat{\mathcal I}^{(b)}_{MT}$'s from the $NP$ permuted samples that exceed $\hat {\mathcal I}_{MT}$.   The test is implemented in  the R-package \textsc{steadyICA} with a default $NP$ value of 199.

We reject independence of the components in $S$ if  the  $p-$value  is less than a prescribed  nominal size.  In principle, the null hypothesis of independence can be rejected because  $u_t$ is not independent, or because the identifying restrictions are incorrect, or both.  But under the maintained assumption that the components of $u$ are mutually independent, the test provides a validation of the (overidentified or exactly identified)  restrictions on $B$ (or $W$).

\section{Simulations}

 The dynamic effects of a disaster shock can be analyzed as follows. Step 1 estimates the coefficients of a VAR model using least squares. Step 2 prewhitens the VAR residuals. Step 3 applies ICA to obtain independent components and associates the component with the largest kurtosis as the disaster shock. Step 4 estimates the impulse response functions. 
Their dynamic effects  after $h$ periods  defined by
 $\Psi(L) =(I-A_1L-\ldots A_PL^p)^{-1}B$     can be computed once consistent estimates of $A$ and $B$ are available. We are primarily interested in  the effects of  $u_{t1}$  on $Y_{t+h}$ and can also estimate the first column of $\Psi^{(h)}$  by projecting the response variable of interest   on $\hat u_1$ on other controls as in \citet{jorda-05}. However, it should be noted that the coefficient estimates from the local projection regressions will have non-standard properties in view of Proposition \ref{prop:prop1}.

To illustrate the effectiveness of this methodology, 
simulations are performed with  $(T,n)=(400,3)$ for
four SVAR(1)  models   based on  two specifications of $B$ and two sets of primitive shocks, holding the $A$ matrix fixed throughout at 
\[ A= \begin{pmatrix} 0.2 & 0 & 0 \\
0.3 & 0.6 & 0 \\
0.4 & 0.3 & 0.8 \end{pmatrix}.
\]

\paragraph{The $B$ matrix}
In model 1 (labeled NLT), the $B$ matrix is Not Lower Triangular. In model 2 (labeled LT), $B$ is Lower Triangular. 
\[
B_{\text{init}}(\text{NLT})=\begin{pmatrix} 1 & 3 & 0 \\ 1 & 1 & 0 \\ 0 & 0 & 1 \end{pmatrix} \quad\quad
B_{\text{init}}(\text{LT})=\begin{pmatrix} 1 & 0 & 0 \\ 1.5 & 1 & 0 \\ 2 & 0.5 & 1 \end{pmatrix}
\]
From a given $B_{\text{init}}$ that is either NLT or LT,  its inverse yields a non-normalized $W_{\text{init}}$  from which a normalized $W$  is formed by imposing the constraint that each row sums to one. Then  $B=W^{-1}$ is  used to simulate data and subsequently estimated.  
\paragraph{Innovations $u$}
The first innovation specification (denoted  HL) has one heavy-tailed shock while in  the second  specification (denoted LL), all three shocks have light tails.  In both cases, the shocks are ordered such that $u_1$ has the largest kurtosis and $u_3$ has the smallest.

\[ \text{HL}:
 \begin{cases}\text{rstable}(1.1,0) \\
   \text{t}_5\\
 \text{t}_{10}\end{cases} \quad\quad \text{LL}:
   \begin{cases} \text{pearson}(0,1,2,20) \\
  \text{pearson}(0,1,-2,10)\\
 t_{15}\end{cases}.\]

\paragraph{Prewritening:} 
Let $e^{0}=(e_{1},e_{2},e_{3})$ and  $e^1=(e_{2},e_{3},e_{1})$ denote two assumed orderings with estimated covariances 
  $\widehat{\cov}(e^0)$ and  $\widehat{\cov}(e^1)$ based on  samples from each vector, respectively.
\begin{itemize}
\item[i.] $\tilde e^0= e^0 P_0^{-1}$, $P_0=\text{chol}(\widehat\cov(e^0))$.
\item[ii.] $\tilde e^1 =e^1 P_1^{-1}$,$P_1=\text{chol}(\widehat\cov(e^1))$.
\item[iii.] $\tilde e^2=e^0P_{\text{svd}}^{-1} $, $P_{\text{svd}}=UD^{1/2}U'$, $\text{svd}(\widehat{\cov}(e^0))=UDU'$.
\item[iv.] $\tilde e^{3}=e^0V_0D_0^{-1}$,  $\text{svd}(e^0_t-\bar e^0_t)=U_0D_0V_0'$.
\end{itemize}

\begin{table}[ht]
\caption{Monte Carlo Simulations}
\label{tbl:table1}
\begin{center} 

Panel A:  Permutation Test: $e$ observed \\
Fraction of  $p$ values $\le 0.1$ \\

\begin{tabular}{rr|ccc|cccc|cccc}
\toprule
Model & noise & $u$ & $u^2$  & $e^0$  & $\tilde e^0$ & $\tilde e^1$ & $\tilde e^2$ & $\tilde e^3$ & $\hat u(\tilde e^0)$ & $\hat u(\tilde e^1)$ & $\hat u(\tilde e^2)$ & $\hat u(\tilde e^3)$\\
\midrule
& & \multicolumn{3}{c|}{ observed} &\multicolumn{4}{c|}{Pewhitened $e$} &
\multicolumn{4}{c}{Estimated Shocks $\tilde e \hat W^\prime $}\\
\midrule
NLT & HL & 0.091 & 0.091 & 1 & 0.892 & 0.271 & 1.000 & 0.804 &0.001 & 0.001 & 0.000 & 0.006\\
LT & HL & 0.091 & 0.091 & 1 & 0.008 & 0.972 & 1.000 & 0.864 & 0.002 & 0.145 & 0.027 & 0.047\\
NLT & LL & 0.092 & 0.089 & 1 & 1.000 & 1.000 & 0.902 & 1.000 & 0.000 & 0.000 & 0.000 & 0.001\\
LT & LL & 0.092 & 0.089 & 1 & 0.003 & 1.000 & 1.000 & 1.000 & 0.000 & 0.001 & 0.000 & 0.007\\

\midrule
\end{tabular}

\begin{center}

Panel B: Permutation Test: $ e$ estimated from VAR\\

Fraction of  $p$ values $\le 0.1$ \\

\begin{tabular}{ll|cccc|cccccc}
\toprule
Model & Noise &  $ \hat e^0$ &  $\hat e^1$ & $\hat e^2 $ & $\hat e^3$ & $\hat u(\hat e^0)$ & $ \hat u(\hat e^1)$& $\hat u(\hat e^2)$ & $\hat u(\hat e^3)$ \\ \hline
& & \multicolumn{4}{c|}{Prewhitened VAR residuals} & \multicolumn{4}{c}{Estimated Shocks $\hat e \hat W^\prime$} \\ \hline
NLT & HL & 0.915 & 0.255 & 1.000 & 0.834 & 0.002 & 0.003 & 0.003 & 0.006\\
LT & HL & 0.006 & 0.982 & 1.000 & 0.869 & 0.003 & 0.139 & 0.027 & 0.058\\
NLT & LL & 1.000 & 1.000 & 0.858 & 1.000 & 0.001 & 0.002 & 0.001 & 0.002\\
LT & LL & 0.005 & 1.000 & 1.000 & 1.000 & 0.001 & 0.001 & 0.001 & 0.004\\

\hline
\end{tabular}
\end{center}

Panel C: Amari Distance: $(\hat B, B)$

\begin{tabular}{rr|cccc|cccccc}
\toprule
Model & noise &    $\hat B(\tilde e^0)$ & $\hat B(\tilde e^1)$ & $\hat B(\tilde e^2)$ & $\hat B(\tilde e^3)$ &
 $\hat B(\hat  e^0)$ & $\hat B(\hat  e^1)$ & $\hat B(\hat e^2)$ & $\hat B(\hat e^3)$\\ \hline
&& \multicolumn{4}{c|}{$e$ observed} & \multicolumn{4}{c}{$e$ estimated} \\ \hline
NLT & HL & 0.151 & 0.148 & 0.170 & 0.211& 0.086 & 0.087 & 0.108 & 0.133\\
LT & HL  & 0.118 & 0.486 & 0.266 & 0.361& 0.122 & 0.605 & 0.320 & 0.482\\
NLT & LL & 0.101 & 0.102 & 0.101 & 0.103& 0.086 & 0.087 & 0.085 & 0.087\\
LT & LL  & 0.103 & 0.103 & 0.106 & 0.114& 0.101 & 0.102 & 0.103 & 0.106\\

\hline\hline
\end{tabular}
\end{center}

\end{table}

 Table \ref{tbl:table1} reports  the Type I errors of the independence test described in Section \ref{sec:independence}, calculated as the mean occurrence of  $p$-values  less than 0.1 in 1000 replications. The results in the top left  panel assume that $e_t$ is observed. Regardless of the specification for $u_t$ and $B$, the Type I errors associated with $u_t$ or $u^2_t$ are close to the size of the test. However, since the components of $u_t$ are non-Gaussian by construction,  the test  always rejects independence of $e^0_t$.   Recall that   $\tilde e^0_t$ are constructed from a Choleski decomposition of the sample covariance matrix for $e^0_t$.  Independence of $\tilde e^0_t$ is always rejected when data are generated  from Model NLT but is almost never rejected for model LT because $W$ is lower triangular in model LT. The  prewhitened data   $\tilde e^1_t$, $\tilde e^2_t$ and $\tilde e^3_t$ are based on $\hat W$ matrices that differ from $W$  and hence the test also rejects independence. The top right panel shows that  the permutation test does not reject independence of the signals $\hat u(\tilde e)$ recovered by ICA  except in  Model LT-HL when the test rejects with probability 0.145 in the Monte-Carlo, which is slightly oversized.

The above  results assume that  $e^0$ is observed. Next, we replace $e^0$ with   residuals from estimation of a VAR with one lag. ICA is then applied to the  estimated residuals after  prewhitening. Panel  B of  Table \ref{tbl:table1} shows that the rejection probabilities of the permutation  test are not affected by having to estimate $A$ and $B$ by least squares. As in the case when $e$ is observed, the permutation test cannot reject independence of the primitive shocks identified by ICA  except in the LT-HL case when the rejection probability is 0.139.

A metric for comparing  matrices  is  Amari distance which, for two $p\times p$ matrices  $A_0$ and $A$  with $r_{ij}=[A_0 A^{-1}]_{ij}$, is defined in \citet{bach-jordan:01} as
\[ d(A_0,A)=\frac{1}{2p}\sum_{i=1}^p \bigg(\frac{\sum_{j=1}^p |r_{ij}|}{\max_j |r_{ij||}}-1 \bigg)+
\frac{1}{2p}\sum_{j=1}^p \bigg(\frac{\sum_{i=1}^p |r_{ij}|}{\max_i |r_{ij||}}-1 \bigg).
\]
Though  ICA studies usually report the Amari distance for the unmixing matrix $W$,  the matrix $B$ is of more interest in SVAR since it is gives impact response of the shocks.  We compare the absolute value of two matrices to ensure that differences are not due to a sign flip that is difficult to control in simulations. Panel C of Table \ref{tbl:table1} shows that  all prewhitening methods give similar  Amari distances  except in Model LT-HL  when using $\hat e_0$   gives noticeably smaller errors.  

The results suggest that the method of prewhitening matters, but only in  the LT-HL case,  and there are two possible explanations. One is that in  the LT-HL case the true $B$ (hence $W$) is lower triangular, and  when this  structure is accompanied by a heavy-tailed shock,  much  can be learned from a kurtosis ordering of the VAR residuals. Prewhitening without using this information is inefficient. The second explanation is that as seen from Panel B,  independence  of  $e^0$ cannot be rejected. This suggests that it is desirable to use  prewhitened data that are as close to independent as possible for ICA estimation. Comparing the $p$ value of the permutation test applied to  different sets of prewhitened data can be  useful in this regard. 

The  results thus favor   prewhitening the VAR residuals by Choleski decomposition ordered by  kurtosis.  A closer look   finds that the $A$ and $B$  matrices are precisely estimated using $\hat e_0$  as prewhitened data.  Even without imposing a lower triangular structure, the pattern is recovered precisely whether or not the innovations have heavy tails. The difference  compared to Choleski decomposition is that ICA lets the data speak as to whether the upper triangular entries of $B$ are zero. If the lower triangular structure is true, $Y_1$ is exogenous and one can alternatively estimate the the dynamic causal effects from a regression of $Y_2$ on $Y_1$ and lags of $Y_1, Y_2, Y_3$.

\begin{center}

Mean Estimates of $A$ and $B$ 

\begin{tabular}{lcccc}
Model/Noise  &  True A &    HL &  LL\\ \hline
   
NLT & $\begin{bmatrix*}[r]
0.2 & 0.0 & 0.0\\
0.3 & 0.6 & 0.0\\
0.4 & 0.3 & 0.8\\

\end{bmatrix*}$ &
$\begin{bmatrix*}[r]
0.192 & 0.002 & -0.007\\
0.293 & 0.602 & -0.007\\
0.400 & 0.300 & 0.800\\

\end{bmatrix*} $ &

$\begin{bmatrix*}[r]

0.193 & 0.000 & -0.007\\
0.298 & 0.598 & -0.004\\
0.399 & 0.303 & 0.798\\

\end{bmatrix*}$ 
\\ \\
LT & $\begin{bmatrix*}[r]
0.2 & 0.0 & 0.0\\
0.3 & 0.6 & 0.0\\
0.4 & 0.3 & 0.8\\

\end{bmatrix*}$ &
$\begin{bmatrix*}[r]
0.190 & 0.003 & -0.009\\
0.296 & 0.598 & -0.005\\
0.399 & 0.301 & 0.798\\

\end{bmatrix*}$
& 
$\begin{bmatrix*}[r]

0.195 & 0.001 & -0.003\\
0.303 & 0.593 & -0.002\\
0.401 & 0.303 & 0.792\\

\end{bmatrix*}$
\\
\hline
\end{tabular}
\end{center}

\begin{center}
\begin{tabular}{lccccc}
Model/Noise &  True B &   HL &   LL\\ \hline
NLT & 
$ \begin{bmatrix*}[r]
1.581 & 2.121 & 0\\
1.581 & 0.707 & 0\\
0.000 & 0.000 & 1\\

\end{bmatrix*}$    &

$\begin{bmatrix*}[r]

1.582 & 2.091 & 0.005\\
1.582 & 0.693 & -0.001\\
0.000 & -0.002 & 0.987\\

\end{bmatrix*}$  &

$\begin{bmatrix*}[r]
1.584 & 2.104 & 0.005\\
1.558 & 0.724 & 0.003\\
-0.002 & -0.001 & 1.000\\

\end{bmatrix*}$ 
\medskip \\
LT& $\begin{bmatrix*}[r]
1.00 & 0.000 & 0.000\\
0.75 & 1.250 & 0.000\\
1.00 & 0.208 & 1.339\\

\end{bmatrix*}$ &

$\begin{bmatrix*}[r]

0.999 & -0.003 & -0.005\\
0.749 & 1.231 & 0.015\\
0.999 & 0.202 & 1.319\\

\end{bmatrix*}$ &

$\begin{bmatrix*}[r]
0.985 & 0.017 & -0.001\\
0.727 & 1.240 & 0.002\\
0.976 & 0.220 & 1.334\\

\end{bmatrix*}$ 
\\ 
\hline
\end{tabular}
\end{center}

\section{Applications}

We consider three applications. The first aims to show that the  validity of ordering used in Choleski  can be tested, as suggested by Lemma \ref{lem:IDtest}. The second application estimates an  HL model to shed light on   the dynamic effects of a disaster shock. In the third, HL regressions are used to purge the variations due to \textsc{covid-19} from the data.

\subsection{Example 1: Uncertainty}
Economic theory is inconclusive as to whether episodes of heightened  uncertainty during economic downturns arise because of  exogenous increases in uncertainty, or if they are the consequence of endogenous responses to other economic shocks.  SVARs have been estimated using a variety of identification strategies  using different measures of uncertainty and over different samples. But testing the validity of these restrictions has been difficult as these models are often  exactly identified, i.e., the number of unique entries in the covariance matrix for $e_t$  equals the number of free parameters in $B$.   An independence test provides a way to test these restrictions.

We  take industrial production (IP) as indicator of  real activity and consider six different measures of uncertainty used  in \citet{lmn1}. These are JLN  macro uncertainty (UM), real economic uncertainty (UR),  financial uncertainty (UF), policy uncertainty (EPU), news-based uncertainty (EPN), and stock market volatility (VIX). This leads to estimation of six  three-variable SVARs,  each using six lags, over the sample 1960:7-2015:4.  Table \ref{tbl:ptest} shows that  the data used in  the six systems have  different statistical properties.  However,  there is little evidence that  the systems considered have heavy tails. 

\begin{table}[ht]
\caption{Permutation Test for $\hat u$ in Three Variable VARs}
\label{tbl:ptest}
\begin{center}
\begin{tabular}{l|ccccccc} 
& \multicolumn{6}{c}{Model}\\ 
 & 1 & 2 & 3 & 4 & 5 & 6\\\hline
variables  & (um,ip,uf) & (ip,ur,uf) & (ip,epu,uf) & (ip,epn,uf) & (ip,epu,vix) & (ip,ur,vix)\\
\hline
ordering & \multicolumn{6}{c}{$p$ values of independence test} \\ \hline
1,2,3 & 0.025 & 0.005 & 0.120 & 0.130 & 0.065 & 0.100  \\
2,1,3 & 0.040 & 0.025 & 0.080 & 0.105 & 0.065 & 0.120  \\
3,1,2 & 0.295 & 0.020 & 0.095 & 0.085 & 0.110 & 0.245  \\
2,3,1 & 0.445 & 0.040 & 0.180 & 0.125 & 0.050 & 0.225  \\
3,2,1 & 0.320 & 0.050 & 0.140 & 0.080 & 0.130 & 0.375  \\
1,3,2 & 0.035 & 0.015 & 0.150 & 0.065 & 0.090 & 0.190\\
\hline
& \multicolumn{6}{c}{kurtosis} \\ \hline
$\kappa_4$: $Y$ 

    & 6.155 & 5.206 & 9.851 & 9.851 & 9.139 & 9.139\\    
    & 5.206 & 4.886 & 3.720 & 5.473 & 7.346 & 9.113\\    
    & 3.436 & 3.436 & 3.421 & 3.421 & 3.645 & 7.346\\    \hline
$\kappa_4$: $\hat e$

    & 4.897 & 4.572 & 6.647 & 6.562 & 7.143 & 5.991\\    
    & 4.575 & 6.509 & 7.209 & 8.770 & 7.591 & 5.474\\    
    & 21.797 & 21.889 & 31.928 & 31.749 & 8.512 & 6.437\\
\hline
$T$ & 652 & 652 & 358 & 358 & 297 & 297\\ \hline

\end{tabular}
\end{center}
Note: \textsc{ip} is industrial production. Six measures of uncertainty are considered: macro (\textsc{um}),  financial (\textsc{uf}),  policy uncertainty (\textsc{epu}), news uncertainty (\textsc{epn}), and stock market volatility index (\textsc{vix}). See  \citet{jln:15} and \citet{lmn1} for definitions.

\end{table}

We  test independence of the identified shocks  obtained  from different orderings of the VAR residuals.  Recall that  the $p$ value indicates  the Type 1 error in rejecting the assumed lower triangular structure. The $p$ values reported in  Table \ref{tbl:ptest} indicate
strong  evidence  against independence of  the shocks constructed from Models  2,4,5  regardless of ordering. There is  some  support for independence when  financial uncertainty is ordered first in Models 1 and 6, while the strongest evidence for independence is provided by  Model 1  using the  ordering (ip,uf,um), a configuration that would not be obvious based on  economic reasoning.

 As Lemma \ref{lem:IDtest} indicates,
 independence is  necessary but not sufficient  for model identification. Nonetheless, testing independence of $\hat u$  provides a way  to rule out incorrect  restrictions. The finding that independence of  shocks from  multiple orderings cannot be rejected  suggests that the restrictions imposed by the Choleski orderings are not enough to uniquely identify  $u$.  This   lends support to  using  restrictions  beyond  the ordering of variables to help identification.

\subsection{Example 2: Disaster Shocks}
The second example considers a SVAR in the cost of disasters series (CD) shown in Figure \ref{fig:cddd}, unemployment claims (Claims), and JLN uncertainty (UM) for the sample 1980:1-2019:12. To make the scale of the variables comparable, the CD series (originally in billions of dollars) is divided by 1000; the claims series is  divided by 1000 so that it is in millions;  the UM series is multiplied by 1000 and remains unit free.  Each series is passed through the filter proposed in \citet{mueller-watson:17} to remove the low frequency variations in the mean. This is equivalent to adding a set of cosine predictors  in the VAR.   The residuals from  estimating a VAR  with six lags  are mean zero with standard deviation  (13.617, 14.276, 8.783) and kurtosis  (69.122, 5.338, 5.232) respectively. The permutation test cannot  reject the null hypothesis of independence of the shocks obtained by Choleski decomposition for orderings (1,2,3), (2,1,3), and (1,3,2). However, irrespective of the ordering of $\hat e$,  the ICA shocks  always pass the independence test. Furthermore,  the shocks identified by the different orderings have very similar   kurtosis.

The ICA estimates obtained  with $\hat e_0$ as  prewhitened  data are:
\begin{eqnarray*}
\hat A_1 &=& \begin{bmatrix*}[r]
0.181* & 0.196  & 0.088*  \\ 
0.049* & 0.911*  & 0.026*  \\ 
0.101* & 0.229  & 1.647*  \end{bmatrix*} 
\quad\quad
\hat A(1) = \begin{bmatrix*}[r]
0.034 & -0.005  & -0.006  \\ 
-0.156  & 0.827*  & 0.136* \\ 
-0.090  & -0.014 & 0.938*  \end{bmatrix*} 
\end{eqnarray*}
According to Proposition \ref{prop:prop1},  $\hat A$ is consistent, though the entries  have different convergence rates. Since the estimates have non-standard distribution, we  use (*) to indicate that zero is outside the (10, 90) percentiles of the bootstrap distribution. The matrix $A_1$ gives the lag one response to a disaster shock.     The   estimates indicate that  response of CD and UM  are both non-zero. The $A_{13}$ estimate suggests  that  the costly disaster series  is not strictly exogenous.  The matrix $\hat A(1)=\sum_{j=1}^p \hat A_j$ summarizes the cumulative effects of the shocks over six periods. The (1,1)-th diagonal entry of $\hat A(1)$ indicates that  the disaster shock has a short half-life.
 The $B$ matrix  gives the instantaneous effect of the disaster shock. The unconstrained ICA estimate is quite close to the one  implied by Choleski decomposition with a (1,2,3) ordering. Taking sampling uncertainty into account, ICA supports a $B$ matrix that is more sparse than the lower triangular structure imposed by Choleski decomposition. Note also that the   HL model is based on the premise that the effects of an infinite variance shock on a finite variance variable are small. Rows two and three of the first column of $B$   and $A_{k}$ are small relative to the own effect recorded in the (1,1)-th entry of the respective matrices.  The estimates are consistent with the HL structure.

\begin{eqnarray*}
\hat B_{chol}&=&\begin{bmatrix*}[r]
13.884* & 0 & 0\\
-0.089  & 14.556*  & 0\\
1.014* & -0.0005  & 8.899* \\
 \end{bmatrix*} = \begin{bmatrix*}[r] 1.0 & 0 & 0 \\ -0.0064& 1.0 & 0\\ 0.073 & -0.0038 & 1.0 \end{bmatrix*} 
\begin{bmatrix*}[r] 13.884 & 0 & 0 \\ 0 & 14.556 & 0 \\ 0 & 0 & 8.899\end{bmatrix*}\\
 \hat B_{ica}&=&\begin{bmatrix*}[r]
13.881*  & 0.153  & 0.235  \\
-0.265  & 14.522*  &0.952   \\
0.870*   & 0.579  & 8.894* \\ 
\end{bmatrix*}=\begin{bmatrix*}[r] 1.0 & 0.010 & 0.026 \\ -0.019 & 1.0 & 0.107 \\ 0.062 & -0.039 & 1.0 \end{bmatrix*} \begin{bmatrix*} 13.881 & 0 & 0 \\ 0 & 14.522 & 0\\ 0 & 0 & 8.894 \end{bmatrix*}
\end{eqnarray*}

The three  shocks recovered by ICA have kurtosis $(69.34, 5.39, 5.03)$. The density of $\hat u_1$ in  Figure \ref{fig:density} shows that the shock has a heavy right tail.  We estimate the impulse response functions by (i) iterating  $A^hB$ as implied by the VAR, (ii) local projections using $\hat u$ obtained from ICA as shocks, and (iii) dynamic responses as implied by Choleski decomposition. These are labeled \textsc{var}, \textsc{lp}, and \textsc{chol} in Table \ref{tbl:xxx}. To provide some idea of precision of the estimated impulse responses, we report standard errors for the \textsc{lp} estimates as well 95\%  bootstrap confidence intervals using the \textsc{vars} package in \textsc{R} as rough guides.  But note that our estimates $\hat A$  have non-standard distributions and results about bootstrap inference with  heavy-tailed variables have only been considered in the univariate setting, see, for example, \citet{davis1997bootstrapping} and \citet{wan-davis:22}. Each local projections regression is of the HL type and hence inference is also non-standard. The standard errors should be interpreted with this caveat in mind.

The CD series has short memory and the effects of its own shock die out after one month. The shock
induces a tightly estimated increase in uncertainty  for  three months and an increase in  unemployment claims of two months. As a point of reference, an unemployment claims shock has an impact effect of 14.556 on itself, and an uncertainty shock has an impact effect of 8.898 on itself. The effects of a disaster shock on these economic variables are small, but they do exist. This   reinforces the motivation of the HL model that infinite variance shocks can affect variables with finite variances.

It can be argued that the infinite variance nature of $u_{t1}$ makes the unit variance property of shocks identified by ICA unappealing. But it is easy  to calibrate the shock  to yield exactly a  one percent change to the variable of interest\footnote{ The 'unit effect normalization' considered in \citet{stock-watson:macrohandbook2} can be used  we transform the ICA estimates  
$\hat u_{t1} =\hat u_{t1,ica} \hat B_{11,ica}, $
$\hat B_{ii} = 1$, and $
\hat B_{j1}= \hat B_{j1,ica}/\hat B_{11,ica}$ for $ j>1$.
}  without changing the shape of the impulse response function.   With this data, the unit effect is associated with  a shock of size 13.881, which is slightly larger than Katrina shock in 2005, which was of magnitude 11.56.

\begin{table}

\caption{Dynamic Effects of a Costly Disaster (CD) Shock}

\label{tbl:xxx}

\begin{center}

\begin{tabular}{r|rrr|rrrrr}
  \hline\hline
$h$ & var & lp & lp.se & chol & chol.up & chol.dn\\
\hline\hline
& \multicolumn{6}{c}{Response of CD} \\ 
\hline
1  & 13.88 & 13.88 & 0.00 & 13.88 & 18.45 & 7.90 \\ 
  2 & 2.63 & 2.61 & 1.87 & 2.65 & 3.85 & 1.23 \\ 
  3 & -0.15 & -0.31 & 0.82 & -0.14 & 0.96 & -1.24 \\ 
  4 & -0.05 & -0.09 & 0.34 & -0.04 & 0.97 & -0.90 \\ 
  5 & -0.51 & -0.44 & 0.35 & -0.51 & 0.70 & -1.54 \\ 
  6 & -0.71 & -0.56 & 0.32 & -0.71 & 0.51 & -1.63 \\ 
   \hline
& \multicolumn{6}{c}{Response of Unemployment Claims} \\ 
\hline
 1  & -0.27 & -0.27 & 0.00 & -0.09 & 1.32 & -1.49 \\ 
  2 & 2.68  & 2.95  & 1.02 &  2.88 & 4.74 & 0.67 \\ 
  3 & 0.18  & 0.37  & 0.63 & 0.39 & 2.43 & -1.04 \\ 
  4 & -0.83 & -0.72 & 0.83 & -0.61 & 1.41 & -2.85 \\ 
  5 & -1.31 & -1.03 & 1.08 & -1.07 & 1.83 & -3.68 \\ 
  6 & -1.76 & -1.64 & 1.15 & -1.54 & 1.50 & -4.33 \\ 
   \hline
& \multicolumn{6}{c}{Response of JLN Uncertainty} \\ 
\hline
1 & 0.87 & 0.87 & 0.00 & 1.01 & 2.21 & 0.16 \\ 
  2 & 2.65 & 2.57 & 0.58 & 2.89 & 4.95 & 1.36 \\ 
  3 & 2.28 & 2.02 & 0.86 & 2.56 & 5.11 & 0.67 \\ 
  4 & 1.26 & 0.74 & 1.10 & 1.56 & 4.64 & -0.67 \\ 
  5 & 0.82 & 0.13 & 1.17 & 1.12 & 5.12 & -1.78 \\ 
  6 & 0.28 & -0.49 & 1.30 & 0.56 & 4.58 & -2.81 \\ 
\hline
\end{tabular}
\end{center}

Note: \textsc{var} are the dynamic response implied by the VAR using the  ICA estimates. \textsc{lp} are responses estimated by local projections, and \textsc{lp.se} are the corresponding standard errors. \textsc{chol} are the dynamic responses using Choleski decomposition, with 95\% confidence intervals defined by \textsc{chol.up} and \textsc{chol.dn}.

\end{table}

\subsection{Example 3: Economic Shocks}
\textsc{covid-19} has been costly in health, social, and economic dimensions, but it has also created new challenges for data analysis. One problem discussed in \citet{ng-covid:21} is that   \text{covid-19} is  pervasive and persistent, and  the principal components of   economic data will be now spanned by  common economic variations and \textsc{covid-19}.  To isolate the economic factors, a suggestion was made to project each economic variable  on covid indicators such as positivity rate, hospitalization, and deaths, then use the panel of `de-covid' data to estimate the economic  factors.  

\textsc{covid-19} also has implications for   VAR estimation. 
Consider  a two variable VAR in log payroll-employment (PAYEMS) and log consumption of durables (CD).  The top panel of Figure \ref{fig:covid} shows the response to a positive employment shock from  a VAR estimated over the pre-covid sample of 1960:1-2020:2, while the second panel extends the sample to 2020:12. Adding ten months of post-covid data completely changed the shape of the impulse response functions. \citet{lenza-primiceri:20} recognize that the covid-induced spikes in the data will distort  VAR estimation and suggest to use Pareto priors for innovation variances to capture these spikes. Others such as \citet{ccmm:21} model \textsc{covid-19} as outliers.

Instead of  specifying changes to the probability distribution of existing shocks, an alternative is to assume as in \citet{ng-covid:21} that there is an additional `virus' shock, say, $v$ in the post-covid sample. There are then two ways to proceed. The first is  to de-covid all variables used in the VAR which would entail running $n(p+1)$ decovid regressions.      By Frish-Waugh arguments, this is the same as adding  covid indicators as exogenous  variables to each equation. Note that these  are not the same as running a VAR on $n$ de-covid  variables, which would only entail $n$ decovid regressions.   Results using the log changes in positive cases as $v$  are shown in the third panel of Figure \ref{fig:covid}.\footnote{The covid data are taken from \url{https://covidtracking.com/data/download}.}  The dynamic responses are  very similar to the ones in the top panel estimated on the pre-covid sample.

Removing the covid variations from the data before VAR estimation   suppresses  feedback from the economic variables to $v$ which could be restrictive. An alternative approach is to include a $v$ indicator in the VAR  directly and order it first, resulting in a HL model with $(n+1)$ variables. In this case, interest is not in the dynamic effects of an infinite variance shock; but  to isolate  the economic variations so that the dynamic effects of economic shocks can be estimated in spite of the presence of \textsc{covid-19}. The results for the three variable VAR in the bottom panel of Figure \ref{fig:covid} are  again similar to the two step approach in the third  panel.     Whichever way we choose to control for covid variations,  the exercise involves regressions with a finite variance variable on the left hand side and a heavy-tailed variable on the right hand side, and  Proposition \ref{prop:prop1} is relevant to the interpretation of the estimates.

\section{Conclusion}
This paper provides a VAR framework that accommodates  disaster-type events. The  framework can be used to study the effects of disaster type shocks, as well as the effects of  finite variance shocks in the presence of large rare events.  Under the maintained assumption that the primitive shocks are independent, a  disaster-type shock can be uniquely identified from the tail behavior and sign of the components estimated by ICA.  An  independence test for validity of the identifying restriction is also proposed. The test is valid even for exactly identified models and is of interest in its own right. The focus here is developing the HL framework and consistent estimation. Inference when the data have heavy tails remains an area for future research.


\clearpage

\begin{figure}[ht]
\caption{Real Cost of Disasters: 1980:1-2019:12}
\label{fig:cddd}%
\begin{center}

\includegraphics[width=6.0in,height=3.00in]{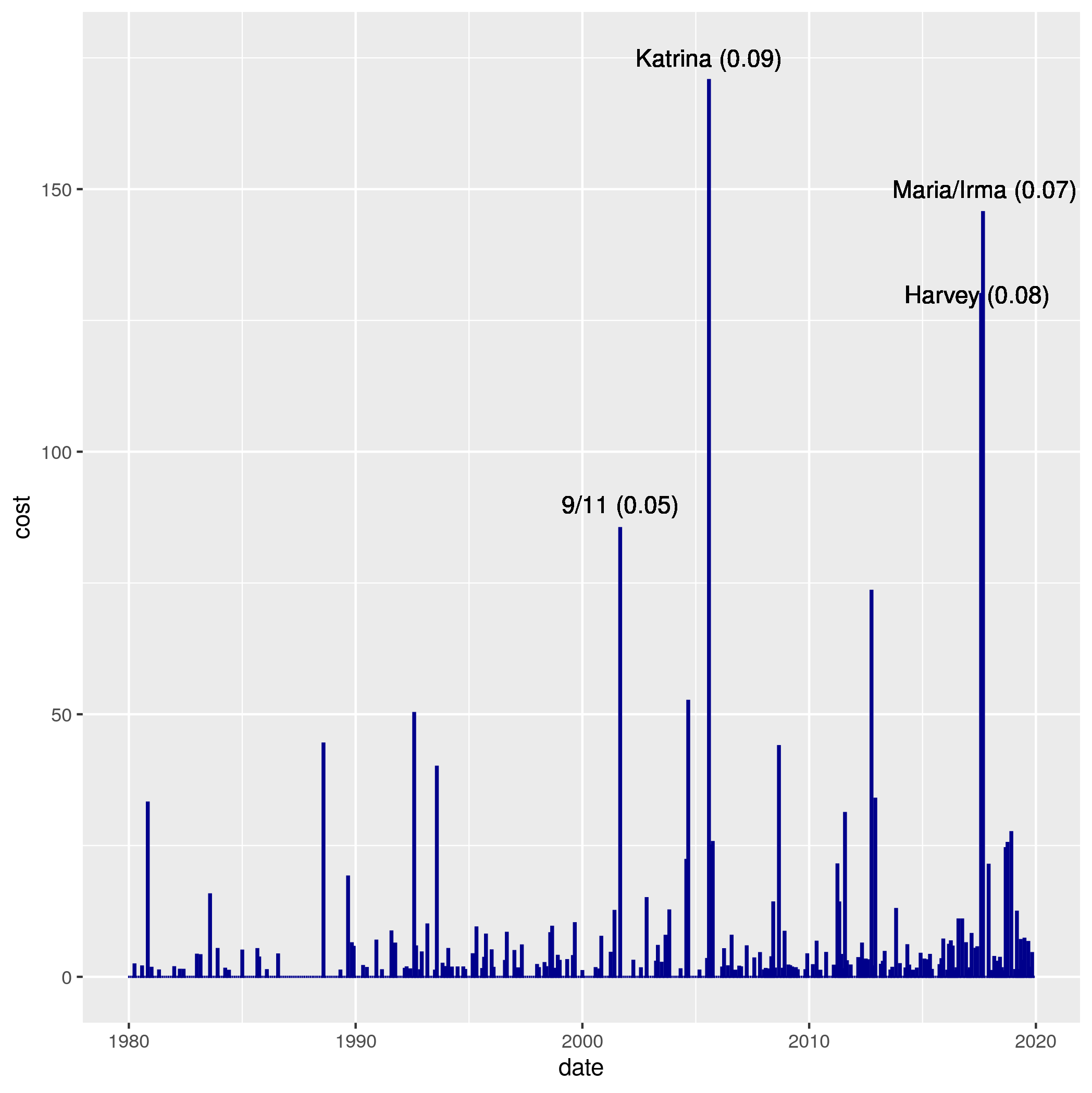}

\end{center}

\end{figure}

\begin{figure}[ht]
\caption{Density of Costly Disaster Shock}
\label{fig:density}
\begin{center}
\includegraphics[width=6.0in,height=2.00in]{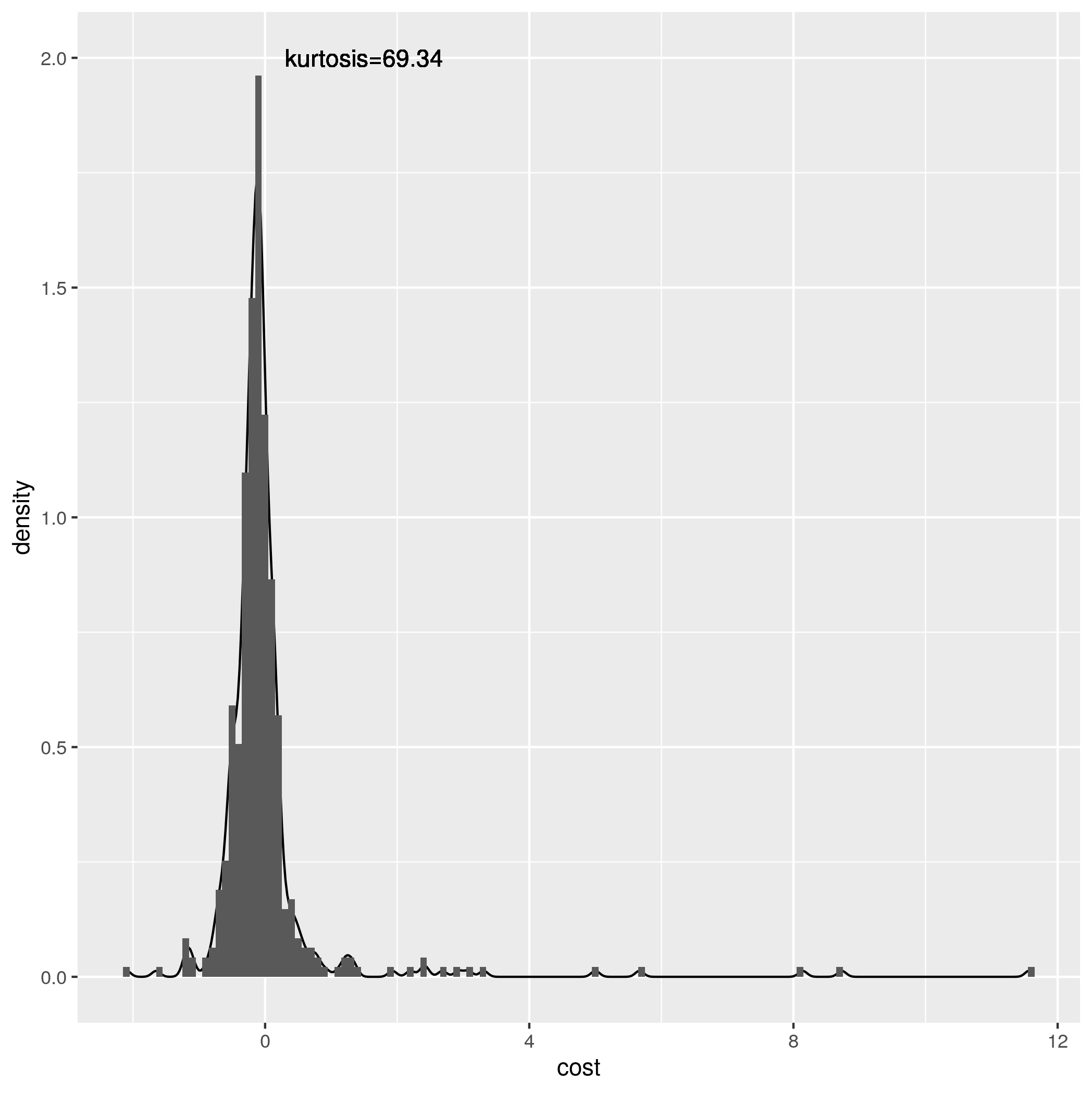}
\end{center}
\end{figure}

\begin{center}
\begin{figure}[ht]
\caption{Dynamic Response of (PAYEMS, CS) to PAYEMS Shock}
\label{fig:covid}
\centering Bivariate VAR, PreCovid

\includegraphics[width=7.0in,height=1.5in]{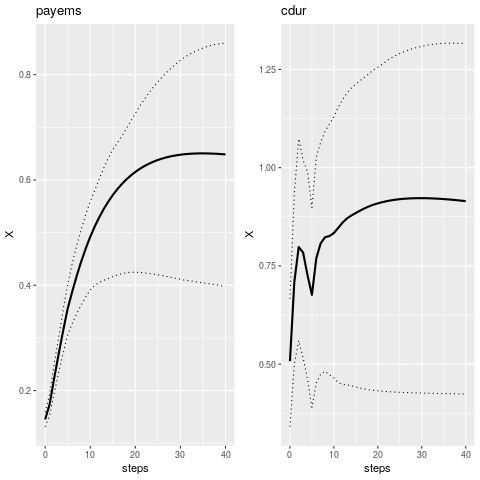}

\centering Bivariate VAR, PostCovid

\includegraphics[width=7.0in,height=1.5in]{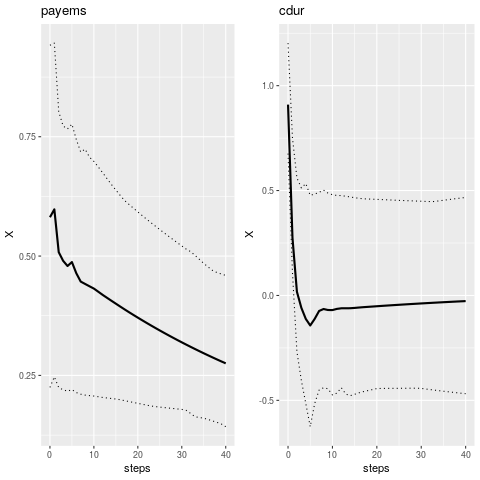}

\centering Bivariate VAR, Purging Covid Effects

\includegraphics[width=7.0in,height=1.5in]{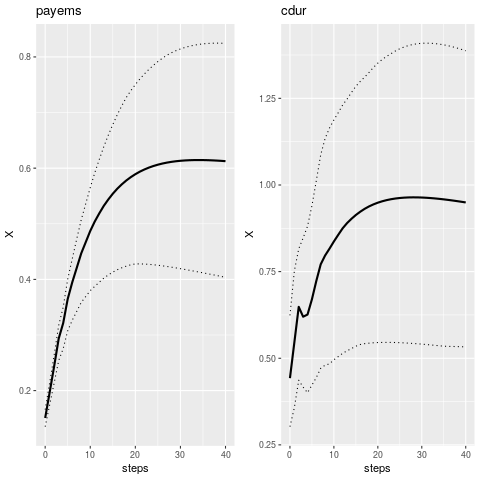}

\centering Three variable VAR, PostCovid

\includegraphics[width=7.0in,height=1.50in]{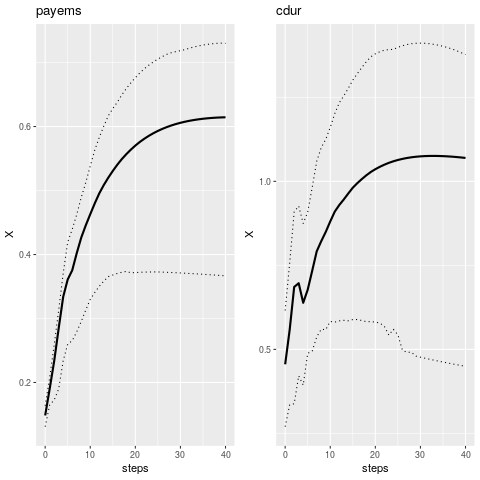}

\end{figure}
\end{center}

\clearpage
\baselineskip=12.0pt
\bibliography{metrics,macro,factor,metrics2,../../../prog1/weekly/notes/weekly}

\section{Appendix
}
\subsection{Background on Distance Covariance}\label{sec:distcova}
The distance covariance between two random vectors $X$ and  $Y$ of dimensions $m$ and $n$, respectively, is given by
\begin{equation}\label{eq:distcova}
{\mathcal I}(X,Y;w)= \int_{\bbr^{m+n}} \big|\varphi_{X,Y}(s,t)-\varphi_X(s)\,\varphi_Y(t)\big|^2 w(s,t)ds\,dt\,,
\end{equation}
where $w(s,t)>0$ is a weight function and $\varphi_Z(t) = E [\exp^{i (t,Z)}],\, t\in\bbr^d \,$ denotes the characteristic function for any random vector
$Z\in\bbr^d$. It is assumed that the integral in \eqref{eq:distcov} is finite, which certainly holds if $w(s,t)$ is a probability density function.  One sees immediately that $X$ and $Y$ are independent if and only if ${\mathcal I}(X,Y;w)=0$ since in this case the joint characteristic function factors into the product of the respective marginal characteristic functions, $\varphi_{X,Y}(s,t)=\varphi_X(s)\,\varphi_Y(t)$ for all $(s,t)\in \bbr^{m+n}$. Now if the weight function factors into a product function, i.e., $w(s,t)=w_1(s)w_2(t)$, then under suitable moment conditions on $X$ and $Y$, ${\mathcal I}(X,Y,w)$ has the form
\begin{eqnarray}
T(X,Y;w)
&=& E[\hat w_1(X-X')\,\hat w_2(Y-Y')] + E[\hat w_1(X-X')]  E[\hat w_2(Y-Y')] \nonumber\\
&& - 2\, E[\hat w_1(X-X')\,\hat w_2(Y-Y'')]\,,
\end{eqnarray}
where $\hat w_1(x)=\int_{\bbr^m}e^{i(s,x)}w_1(s)ds$, $\hat w_2(y)=\int_{\bbr^n}e^{i(t,y)}w_1(s)ds$, and $(X,Y), (X',Y'), (X'',Y'')$ are iid copies of $(X,Y)$. This relation is found by expanding the square in the integrand in \eqref{eq:distcova} and using Fubini to interchange integration with expectation.  Precise conditions on $w$ to perform these operations is given in \citet{davis2018applications}.  Suffice it to say that if $w_i$ is a probability density function, then application of Fubini requires no further conditions on the distributions of $X$ and $Y$.  In order to avoid direct integration in \eqref{eq:distcova}, one can choose functions, $w_i$ which have an easily computable Fourier transform.  Examples include the Gaussian density for which $\tilde w_i(x)=\exp\{-\sigma^2 \|x\|^2/2\}$ or a Cauchy density in which case $\tilde w_i(x)=\exp\{-\sigma \|x\|_1\}$, where $\|x\|_1$ is the 1-norm.  A popular choice for $w$ is
\begin{equation}\label{eq:szekelya}
w(s,t)=\left(c_{m,\beta}|s|^{\beta+m}c_{n,\beta}|t|^{\beta+n}\right)^{-1}\,,
\end{equation}
where $\beta\in (0,2)$,  $c_{m,\beta}=\frac{2\pi{m/2}\Gamma(1-\beta/2)}{\beta2^\beta\Gamma((\beta+m)/2)}$  (see \citet{szekely-rizzo-bakirov:07}).  In this case, one has $\int_\bbr^m c_{m,\beta}^{-1}\left(1-\cos (s,x)\right)ds=|x|^{\beta}$, and provided $E(|X|^\beta+E|Y|^\beta+|X|^\beta|Y|^\beta)<\infty$, then
\begin{eqnarray}\label{eq:distcov2a}
{\mathcal I}(X,Y;w)
&=& E[|X-X'|^\beta|Y-Y'|^\beta)] + E[|X-X'|^\beta]  E[|Y-Y'|^\beta] \nonumber\\
&& - 2\, E[|X-X'|^\beta|Y-Y''|^\beta]\,.
\end{eqnarray}
Notice that with this choice of $w$, ${\mathcal I}$ is invariant under orthogonal transformations on $X$ and $Y$ and is scale homogeneous under positive scaling.  The most common choice for $\beta$ is the value 1, which requires a finite mean.  In our heavy-tailed framework, we have assumed the tail-index $\alpha\in (1,2)$ so the integral in \eqref{eq:distcova} is finite and formula \eqref{eq:distcov2a} is valid (if $X$ and $Y$ are independent)  using the above weight function $w$ with $\beta=1$.  However, in order to extend the results to heavier tails, such as Cauchy, then one can choose a smaller $\beta$, which is difficult to identify in practice, or use the the Gaussian density function.  As noted in \citet{davis2018applications}, the weight function in \eqref{eq:szekely} can have potential limitations when applied to estimated residuals in the finite variance case.

Based on data $(X_1,Y_1),\ldots,(X_T,Y_T)$ from $(X,Y)$, the general distance covariance in \eqref{eq:distcova} can be estimated by replacing the characteristic function with their empirical counterparts. 
Using the  $w$  given in \eqref{eq:szekelya}, we obtain the estimate 
\begin{eqnarray*}\label{eq:distcovhat}
	\hat{\mathcal I}(X,Y;w)&=&\frac{1}{T^2}\sum_{i=1}^T\sum_{j=1}^T |X_i-X_j|^\beta\,|Y_i-Y_j|^\beta+
	\left(\frac{1}{T^2}\sum_{i=1}^T\sum_{j=1}^T |X_i-X_j|^\beta\right)\left(\frac{1}{T^2}\sum_{i=1}^T\sum_{j=1}^T |Y_i-Y_j|^\beta\right)\\
	&&~~~-2\frac{1}{T^3}\sum_{i=1}^T\sum_{j=1}^T\sum_{k=1}^T |X_i-X_j|^\beta\,|Y_i-Y_k|^\beta\,,
\end{eqnarray*}
which can be shown to be consistent for ${\mathcal I}(X,Y;w)$ by the ergodic theorem applied to the empirical characteristic function.  The limit theory for $T\hat{\mathcal I}(X,Y;w)$, under the assumption that $X$ and $Y$ are independent can be found in \citet{szekely-rizzo-bakirov:07} in the iid case and in  in \citet{davis2018applications} a time series setting when $\{(X_t,Y_t)\})$ is a stationary time series. The latter also considers the limit theory of $\sqrt{T}\left(\hat{\mathcal I}(X,Y;w)-{\mathcal I}(X,Y;w)\right)$, when $X$ and $Y$ are not independent.

\subsection{Proofs}

We now consider an array of models given by $Y_{t,T}=A_{T}Y_{t-1,T}+B_T u_t$, where 
\begin{itemize}
   \item[(i)]  $A_{21,T}=\frac{a_{21}}{T^{\theta}}$ and $B_{21,T}=\frac{b_{21}}{T^{\theta}}$, with $\theta=1/\alpha-1/2$.
\item[(ii)] $u_{t1}$ has Pareto-like tails, $\mathbb E[u_{t1}]=0$ if it exists and has dispersion 1 so that $T\mathbb P(|u_{t1}|>T^{1/\alpha})\rightarrow 1$ and $u_{t2}\sim (0,1)$.
\end{itemize}
In other words, for fixed $T$, the time series $\{Y_{t,T},\,t\in \mathbb Z\}$ satisfies the VAR(1) equations with coefficient matrix $A_T$.  The time series of observations, $Y_1,\ldots,Y_T$ are then considered to come from this model. To lighten the notation going forward, we will often suppress the dependence of $Y_t$ on $T$.
\paragraph{Claim 1. }  We will first show  that the assumptions  imply that 
$ \frac{1}{T} \sum Y_{t2,T}^2\dconv S^*_2$ where $S^*_2$ is random or a constant, and hence the sample variance of $Y_{t2}$ is convergent.

From the causal MA representation, we have
\begin{eqnarray*}
Y_t=\sum_{j=0}^\infty A_T^jBu_{t-j}\,,	
\end{eqnarray*}
which can be expressed component-wise as  
\begin{eqnarray}
Y_{t1}&=&\sum_{j=0}^{\infty}c_{j,11}u_{t-j,1}+\sum_{j=0}^{\infty}c_{j,12}u_{t-j,2}\label{eq:yt11}\\
Y_{t2}&=&\frac{1}{T^\theta}\sum_{j=0}^{\infty}c_{j,21}u_{t-j,1}+\sum_{j=0}^{\infty}c_{j,22}u_{t-j,2}\,,\label{eq:yt21}
\end{eqnarray}	
where  the $c_{j,ik}$   depend on $T$ but converge (as $T\to \infty$) to finite limits that decay as a function of the lag $j$ at an exponential rate.  For example, $(c_{0,11},c_{0,12})$ corresponds to the first row of the matrix $B$, while $(c_{0,21},c_{0,22})=(b_{21},B_{22})$.



Since  $\theta=1/\alpha-1/2$,   $2\theta+1=2/\alpha$, and $1+\theta=1/\alpha+1/2$, 
the sample mean of the $Y_{t2}$, normalized by $\sqrt T$,  converges in distribution even if the population variance is infinite.  We have
\begin{eqnarray*}
 \frac{1}{\sqrt{T}} \sum_{t=1}^T Y_{t2}&=&\sum_{j=0}^{\infty}c_{j,21} \left(\frac{1}{T^{1/\alpha}}\sum_{t=1}^Tu_{t-j,1}\right) +\sum_{j=0}^{\infty}c_{j,22} \left(\frac{1}{\sqrt{T}}\sum_{t=1}^Tu_{t-j,2}\right) \\
&\dconv &\left(\sum_{j=0}^{\infty}c_{j,12}^*\right)S_{u}+\left(\sum_{j=0}^{\infty}c_{j,22}^*\right)N_{u}=:S^*_{Y,2}
\end{eqnarray*}
where $c_{j,2i}^*=\lim_{T\to\infty}c_{j,2i}$, $i=1,2$, and $S_u$ is a stable random variable with index $\alpha$ that is independent of $N_u$, a standard normal random variable. 
The sample second moment also converges.  To see this, we have
 \begin{eqnarray}
  	\frac{1}{T} \sum_{t=1}^T Y_{t2}^2 &=& \sum_{j=0}^{\infty}c_{j,21}^2\left(\frac{1}{T^{2/\alpha}}\sum_{t=1}^Tu_{t-j,1}^2\right) + 
  	\sum_{j\ne k=0}^{\infty}c_{j,21}c_{k,21}\left(\frac{1}{T^{2/\alpha}}\sum_{t=1}^Tu_{t-j,1}u_{t-k,1}\right)\nonumber\\
 	&&+\sum_{j=0}^{\infty}c_{j,22}^2\left(\frac{1}{T}\sum_{t=1}^Tu_{t-j,2}^2\right) +	\sum_{j\ne k=0}^{\infty}c_{j,22}c_{k,22}\left(\frac{1}{T}\sum_{t=1}^Tu_{t-j,2}u_{t-k,2}\right)\nonumber\\
 	&&+	\sum_{j, k=0}^{\infty}c_{j,21}c_{k,22}\left(\frac{1}{T^{1/\alpha+1/2}}\sum_{t=1}^Tu_{t-j,1}u_{t-k,2}\right)\nonumber\\
 	&\dconv& \left(\sum_{j=0}^{\infty}(c_{j,21}^*)^2\right)S_{uu} + \sum_{j=0}^{\infty}(c_{j,22}^*)^2 =: S^*_{YY,2}\,,\label{eq:y22} \end{eqnarray}
where $S_{uu}$ is a stable random variable with index $\alpha/2$.  The last line follows essentially from Lemma \ref{lem:lemma1}, which shows that for $j\ne k$, 
\begin{eqnarray}
&&\frac{1}{T^{1/\alpha}\log T}\sum_{t=1}^Tu_{t-j,1}u_{t-k,1}=O_p(1)\label{eq:sum1}\\ &&\frac{1}{T^{1/\alpha}}\sum_{t=1}^Tu_{t-j,1}u_{t-k,2}=O_p(1),\label{eq:sum2} \\ &&\frac{1}{T}\sum_{t=1}^Tu_{t-j,2}u_{t-k,2}=o_p(1)\label{eq:sum3}
\end{eqnarray} 
by the ergodic theorem.  Using the ideas in \citet{davis-resnick:86} and the continuous mapping theorem, it is straightforward to obtain the limit in the last line upon summing out $j$ and $k$.

\subsection*{Proof of Proposition \ref{prop:prop1}}

We first note that the OLS estimate of $A$ is given by 
\begin{equation}\label{eq:ahatols}
 \hat{A}=\sum_{t=2}^TY_{t}Y_{t-1}^\prime \left(\sum_{t=1}^{T-1}Y_tY_t^\prime\right)^{-1} 
 \end{equation}
 and hence
 \begin{equation}\label{eq:ahatm}
 \hat{A}-A=\sum_{t=2}^Te_{t}Y_{t-1}^\prime \left(\sum_{t=1}^{T-1}Y_tY_t^\prime\right)^{-1}
 \end{equation}
(For simplicity, we have terminated the second sum in \eqref{eq:ahatols} and \eqref{eq:ahatm} at $T-1$ instead of $T$, but this has no bearing on the asymptotics.)
We begin by analyzing the terms in the matrix of cross products given by $\sum_{t=1}^{T-1}Y_tY_t^\prime$.
From \eqref{eq:y22}, we can write 
$$
\sum_{t=1}^TY_{t2}^2=TS_{22,T},
$$ where $S_{22,T}=\frac{1}{T}\sum_{t=1}^TY_{t2}^2\dconv S^*_{YY,2}$.
The same argument as above shows that $T^{-2/\alpha} \sum_{t=1}^T Y_{t1}^2\dconv S^*_{YY,1}$, so that  
$$
T^{2/\alpha} \sum_{t=1}^T Y_{t1}^2=T^{2/\alpha}S_{11,T},
$$ with $S_{11,T}\dconv S^*_{YY,1}$.
Now applying the same arguments as before,
\begin{eqnarray*}
\frac{1}{T^{1/\alpha+1/2}}\sum_{t=1}^T Y_{t1} Y_{t2}&=&\sum_{j, k=0}^{\infty}c_{j,11}c_{k,21}\left(\frac{1}{T^{2/\alpha}}\sum_{t=1}^Tu_{t-j,1}u_{t-k,1}\right)
+\sum_{j, k=0}^{\infty}c_{j,12}c_{k,21}\left(\frac{1}{T^{2/\alpha}}\sum_{t=1}^Tu_{t-j,2}u_{t-k,1}\right)\\
&&+\sum_{j, k=0}^{\infty}c_{j,11}c_{k,22}\left(\frac{1}{T^{1/\alpha+1/2}}\sum_{t=1}^Tu_{t-j,1}u_{t-k,2}\right)\\
&&+\sum_{j, k=0}^{\infty}c_{j,12}c_{k,22}\left(\frac{1}{T^{1/\alpha+1/2}}\sum_{t=1}^Tu_{t-j,2}u_{t-k,2}\right)\\
&\dconv& \left(\sum_{j=0}^{\infty}c_{j,11}^*c_{j,21}^*\right)S_{uu} =:  S^*_{YY,12}\,,\\		
\end{eqnarray*}
where we have made use of \eqref{eq:sum1}--\eqref{eq:sum3} once again.  We write 
$$
\sum_{t=1}^T Y_{t1} Y_{t2}=T^{1/\alpha+1/2}S_{12,T},
$$ where $S_{12,T}\dconv S^*_{YY,12}$.
It follows that 
\begin{eqnarray*}
(\sum_{t=1}^TY_tY_t^\prime)^{-1}=\frac{1}{\text{det}}\begin{pmatrix} \sum_{t=1}^T Y_{t2}^2 & -\sum_{t=1}^T Y_{t1} Y_{t2} \\
-\sum_{t=1}^T  Y_{t1} Y_{t-1,2} & \sum_{t=1}^T Y_{t1}^2\end{pmatrix}
=\frac{1}{\text{det}}\begin{pmatrix}TS_{22,T}  & -T^{1/\alpha+1/2} S_{12,T} \\
	-T^{1/\alpha+1/2} S_{12,T} & T^{2/\alpha}S_{11,T}\end{pmatrix}\,,
\end{eqnarray*}
where $det=T^{2/\alpha+1}D_T$ and $D_T=(S_{11,T}S_{22,T}-S_{12,T}^2)\dconv D_\infty:=S_{YY,1}^*S_{YY,2}^*-(S_{YY,12}^*)^2$. The inverse can then be written more concisely as
\begin{eqnarray}\label{eq:yyinv}
	(\sum_{t=1}^TY_tY_t^\prime)^{-1}
	=\frac{1}{D_T}\begin{pmatrix}T^{-2/\alpha}S_{22,T}  & -T^{-1/\alpha-1/2} S_{12,T} \\
		-T^{-1/\alpha-1/2} S_{12,T} & T^{-1}S_{11,T}\end{pmatrix}\,.
\end{eqnarray} 

\paragraph{Estimation of  the $Y_1$ equation}
We have
\begin{eqnarray}
(T\log T)^{-1/\alpha}\sum_{t=2}^T e_{t1}Y_{t-1,1}&=& \sum_{j=0}^\infty c_{j,11}(T\log T)^{-1/\alpha}\sum_{t=2}^Tu_{t-1-j,1}(u_{t1}B_{11}+u_{t2}B_{12})\nonumber \\
&&+\sum_{j=0}^\infty c_{j,12}(T\log T)^{-1/\alpha}\sum_{t=2}^Tu_{t-1-j,2}(u_{t1}B_{11}+u_{t2}B_{12})\nonumber\\
&=&\sum_{j=0}^\infty c_{j,11}(T\log T)^{-1/\alpha}\sum_{t=2}^Tu_{t-1-j,1}u_{t1}B_{11}+o_p(1)\nonumber\\
&\dconv& S_{eY,11}.\label{eq:y1e1}
\end{eqnarray}

Next
\begin{eqnarray}
T^{-1/\alpha}\sum_{t=2}^T e_{t1}Y_{t-1,2}&=& \sum_{j=0}^\infty c_{j,21} T^{-2/\alpha+1/2}\sum_{t=2}^Tu_{t-1-j,1}(u_{t1}B_{11}+u_{t2}B_{12}) \nonumber\\
&&+\sum_{j=0}^\infty c_{j,22}T^{-1/\alpha}\sum_{t=2}^Tu_{t-1-j,2}(u_{t1}B_{11}+u_{t2}B_{12})\nonumber\\
&=&\sum_{j=0}^\infty c_{j,22}T^{-1/\alpha}\sum_{t=2}^Tu_{t-1-j,2}u_{t1}B_{11}+o_p(1)\nonumber\\
&\dconv& S_{eY,12}.\label{eq:y2e1}
\end{eqnarray}

\paragraph{Estimation of the $Y_2$ equation}
Using the representation in \eqref{eq:yt11}, it is relatively straightforward to show
 \begin{eqnarray*}
T^{-1/\alpha}\sum_{t=2}^T e_{t2}Y_{t-1,1} &=& \bigg(\frac{b_{21}}{T^{\theta+1/\alpha}}\sum_{t=2}^T Y_{t-1,1} u_{t1}+B_{22}T^{-1/\alpha}\sum_{t=2}^T u_{t2}Y_{t-1,1}\bigg)
\dconv  B_{22} S_{Yu,21}=:S_{eY, 21}
\end{eqnarray*}
since $(\log T)^{1/\alpha}/T^\theta\rightarrow 0$. Furthermore, since  $\frac{(T\log T)^{1/\alpha}}{T^{2/\alpha-1/2}}=\frac{(\log T)^{1/\alpha}}{T^\theta}\rightarrow 0$, 
\begin{eqnarray*}
	\frac{1}{\sqrt{T}}\sum_{t=2}^Te_{t2}Y_{t-1,2}&=&\sum_{j=0}^\infty c_{j,21}\frac{1}{T^{1/\alpha}}\sum_{t=2}^Tu_{t-j-1,1} \left(\frac{b_{21}}{T^{\theta}}u_{t1}+B_{22}u_{t2}\right) \\
	&&+\sum_{j=0}^\infty c_{j,22}\frac{1}{T^{1/2}}\sum_{t=2}^Tu_{t-j-1,2}\left(\frac{b_{21}}{T^{\theta}}u_{t1}+B_{22}u_{t2}\right) \\	
	&\dconv&  B_{22} S_{Yu,12}+b_{21}S_{uu,21}+B_{22}N_{u2}=:S_{eY,22}\,,
\end{eqnarray*}
where $S_{Yu,12}, S_{uu,21}$ are stable with index $\alpha$ and $N_{u2}$ is normally distributed.  

Summarizing, we have, using an obvious notation,
\begin{eqnarray*}
\sum_{t=2}^Te_tY_{t-1}^\prime	
=\begin{pmatrix}\sum_{t=2}^Te_{t1}Y_{t-1,1} & \sum_{t=2}^Te_{t1}Y_{t-1,2} \\ \sum_{t=2}^Te_{t2}Y_{t-1,1} & \sum_{t=2}^Te_{t1}Y_{t-1,2}\end{pmatrix}
=\begin{pmatrix}(T\log T)^{1/\alpha}\tilde S_{11,T} &T^{1/\alpha} \tilde S_{12,T} \\ T^{1/\alpha}\tilde S_{21,T} &T^{1/2} \tilde S_{22,T}\end{pmatrix}\,,
\end{eqnarray*}
so that $\hat A-A$ is 
\begin{equation*}
=\frac{1}{D_T}\begin{pmatrix}(T\log T)^{1/\alpha}\tilde S_{11,T} &T^{1/\alpha} \tilde S_{12,T} \\ T^{1/\alpha}\tilde S_{21,T} &T^{1/2} \tilde S_{22,T}\end{pmatrix}
\begin{pmatrix}T^{-2/\alpha}S_{22,T}  & -T^{-1/\alpha-1/2} S_{12,T} \\
	-T^{-1/\alpha-1/2} S_{12,T} & T^{-1}S_{11,T}\end{pmatrix}\,.
\end{equation*}
Since $-1/\alpha+1/2<0$ and $1-1/\alpha-1/2<0$, we conclude
\begin{eqnarray*}
	\sqrt{T} (\hat A_{11}-A_{11}) &\dconv& S_{A,11}:=-S_{eY,12}S_{YY,12}^*/D_\infty\\
	T^{1-1/\alpha}(\hat A_{12}-A_{12})&\dconv& S_{A,12}:=S_{eY,12}S_{YY,11}^*/D_\infty\\
	T^{1/\alpha}(\hat A_{21}-A_{21})  &\dconv & S_{A,21}:=(S_{eY,21}S_{YY,22}^*-S_{eY,22}S_{YY,12}^*)/D_\infty\\
	\sqrt{T} (\hat A_{22}-A_{22}) &\dconv & S_{A,22}:=(S_{eY,22}S_{YY,11}^*-S_{eY,21}S_{YY,12}^*)/D_\infty.
\end{eqnarray*}
$\Box$

\subsection*{Proof of Proposition \ref{prop:propWehat}}
The proof relies on an application of Theorem 3.3 in \citet{davisfernandes2021}, which considers consistency for the unmixing matrix in an ICA model with noise.  Observe that 
\begin{equation}\label{eq:app1}
\hat e_t=Y_t-\hat Y_t= e_t+(A-\hat A)Y_{t-1}= Bu_t+r(T)Y_{t-1}\,,
\end{equation}
where $r(T)=A-\hat A=o_P(1)$.  By the independence of $u_t$ with $Y_{t-1}$ it follows that the components of $E|u_t||Y_{t-1}|<\infty$.  Hence, in order to apply Theorem 3.3, it suffices to show that 
\begin{equation}\label{eq:app2}
\hat\Sigma^{-1}_e-\hat\Sigma^{-1}_{\hat e}\cip 0\,.
\end{equation}
We first show 
\begin{equation}\label{eq:app3}
\hat\Sigma_e-\hat\Sigma_{\hat e}\cip 0\,.
\end{equation}
From \eqref{eq:app1},
\begin{eqnarray*}
\hat\Sigma_e-\hat\Sigma_{\hat e}&=&-(\hat A-A)\hat\Sigma_{y}(\hat A-A)^\prime+(\hat A-A)T^{-1}\sum_{t=1}^TY_{t-1}e_{t}^\prime+T^{-1}\sum_{t=1}^Te_tY_{t-1}^\prime(\hat A-A)^\prime,\\
\end{eqnarray*}
where $\hat\Sigma_y$ is the sample covariance matrix of $Y_{t-1},\,t=1,\ldots,T$.  Using the relations for $\hat A-A$ in Proposition \ref{prop:prop1}, and the calculations leading to the limit in \eqref{eq:yyinv}, it follows that  
$$
(\hat A-A)\hat\Sigma_{y}(\hat A-A)^\prime\cip 0\,.
$$
Similarly, applying \eqref{eq:y1e1}--\eqref{eq:y2e1}, it is straightforward to also show that
$$
(\hat A-A)T^{-1}\sum_{t=1}^TY_{t-1}e_{t}^\prime+T^{-1}\sum_{t=1}^Te_tY_{t-1}^\prime(\hat A-A)^\prime
\cip 0\,,
$$
which proves \eqref{eq:app3}.  To finish the proof, we note the following relations 
\begin{eqnarray}
\frac{1}{T}\sum_{t=1}^Te_{t1}^2&=&T^{2/\alpha-1}B_{11}^2S_{11,T}+B_{12}^2\sigma_2^2 +o_P(1)\label{eq:appe1}\\
\frac{1}{T}\sum_{t=1}^Te_{t2}^2&=&b_{21}^2S_{11,T}+B_{22}^2\sigma_2^2+o_P(1)\label{eq:appe2}\\
\frac{1}{T}\sum_{t=1}^Te_{t1}e_{t2}&=&T^{1/\alpha-1/2}B_{11}b_{21}S_{11,T}+B_{11}B_{22}\sigma_2^2+o_P(1)\,,\label{eq:appe3}
\end{eqnarray}
where $S_{11,T}=T^{-2/\alpha}\sum_{t=1}^Tu_{t1}^2=O_P(1)$, and $\sigma_2^2=\mbox{var}(u_{t2})$.  In view of \eqref{eq:app3}, the exact same relations hold for the corresponding entries of $\hat\Sigma_{\hat e}$.  The determinant of both sample covariance matrices is then of order
$$
|\hat\Sigma_e|=T^{2/\alpha-1}B_{11}^2S_{11,T}B_{22}\sigma_2^2+O_P(T^{1/\alpha-1/2})\,.
$$
Since $\hat\Sigma^{-1}_e=\frac{1}{|\hat\Sigma_e|}\tilde \Sigma_e$, where
$$
\tilde\Sigma_e=\begin{pmatrix} T^{-1}\sum_{t=1}^T e_{t2}^2 & -T^{-1}\sum_{t=1}^T e_{t1}e_{t2}\\
-T^{-1}\sum_{t=1}^T e_{t1}e_{t2} &T^{-1}\sum_{t=1}^T e_{t1}^2\end{pmatrix}\,,
$$
with a similar expression for $\hat\Sigma^{-1}_{\hat e}$, we have
$$
\hat\Sigma^{-1}_e-\hat\Sigma^{-1}_{\hat e}=\left(\frac{1}{|\hat\Sigma_e|}-\frac{1}{|\hat\Sigma_{\hat e}|}\right)\tilde \Sigma_e-\frac{1}{|\hat\Sigma_{\hat e}|}\left(\tilde \Sigma_e-\tilde \Sigma_{\hat e}\right)\,.
$$
The second converges to in probability by \eqref{eq:app2} and the fact that he determinant goes to infinity in probability.  To show the first term converges to 0, since $|\hat\Sigma_e|/|\hat\Sigma_{\hat e}|\cip 1$, it suffices to show that the matrix $\hat \Sigma_e^{-1}$ remains bounded.  But
$$
\hat\Sigma_e^{-1}=W\hat \Sigma_u^{-1}W^\prime\,,
$$
where $\hat\Sigma$ is the sample covariance matrix of $u_1,\ldots,U_t$.  A straight forward calculation shows that 
$$
\hat\Sigma_u^{-1}\cip \begin{pmatrix} 0 & 0\\  0 &\sigma_2^{-2} \end{pmatrix}
$$
and hence $\hat\Sigma_e^{-1}$ is bounded in probability as claimed.  This establishes \eqref{eq:app2} which completes the proof of the proposition. $\Box$

\subsection*{Proof of Lemma \ref{lem:choleski}} \label{app:prooflem3}
 Using \eqref{eq:appe1}, we see that
$$
e_{t1}^c=\frac{e_{t1}}{\sqrt{T^{-1}\sum_{s=1}^Te_{s1}^2}}\sim \frac{e_{t1}}{\sqrt{T^{2/\alpha-1}B_{11}^2S_{11,T}}}\,,
$$
which gives the asserted representation for $e_{t1}^c$.  
Also from \eqref{eq:appe2} and \eqref{eq:appe3}, we find that
$$
c_T\sim \frac{b_{21}}{T^{1/\alpha-1/2}B_{11}}\,,
$$
as $T\to\infty$. It follows from \eqref{eq:appe1}--\eqref{eq:appe3} that
\begin{eqnarray*}
T^{-1}\sum_{s=1}^T(e_{s2}-c_Te_{s1})^2&=&T^{-1}\sum_{s=1}^T
\left(e_{s2}^2-2c_Te_{s1}e_{s2}+c_T^2e_{s2}^2\right)\\
&\sim & B_{22}^2\sigma_2^2\,.
\end{eqnarray*}
Since $c_T\cip 0$, we conclude
$$
e^c_{t2}=\frac{e_{t1}}{\sqrt{T^{-1}\sum_{s=1}^T(e_{s2}-c_Te_{t1})^2}}\sim \frac{e_{t2}-c_Te_{t1}}{|B_{22}|\sigma_2}\sim\frac{B_{22}}{|B_{22}|\sigma_2}u_{t2}\,,
$$
as claimed.  $\Box$

\end{document}